\documentclass[iop]{emulateapj}
\usepackage{graphicx}                                                         

\def\msol{\,{\rm /figM}_\odot}              
\def\apj{Astrophys. J. }
\def\apjs{Astrophys. J., Suppl. Ser. }

\def\pre{Physical Review E.}
\def\aap{A\&A}

\def\ga{\,\hbox{\hbox{$ > $}\kern -0.8em \lower 1.0ex\hbox{$\sim$}}\,}
\def\la{\,\hbox{\hbox{$ < $}\kern -0.8em \lower 1.0ex\hbox{$\sim$}}\,}

\def\gcc{\,{\rm g}\,{\rm cm}^{-3}}
\def\ubar{\tilde U}
\def\sbar{\tilde S}
\def\ucp{\tilde {C_P}}
\def\ucv{\tilde {C_V}}
\def\gad{\nabla_{ad}}
\def\kb{k_B}
\def\ln{{\rm Ln}}
\def\fex{F_{ex}}
\def\mh{m_{H}}
\def\msol{{\rm M}_\odot}
\def\mjup{{\rm M}_{\rm Jup}}

\begin{document}

\title{A new equation of state for dense hydrogen-helium mixtures}

\author{G. Chabrier} 
\affil{
CRAL, Ecole normale sup\'erieure de Lyon, UMR CNRS 5574, 69364 Lyon Cedex 07,  France,\\
School of Physics, University of Exeter, Exeter, EX4 4QL, UK
}

\author{S. Mazevet}
\affil{
Laboratoire Univers et Theories, Universite Paris Diderot, Observatoire de Paris, PSL University, 5 Place Jules Janssen, 92195
Meudon France.Luth, Observatoire de Meudon, France,\\
CEA-DAM-DIF, 91280 Bruyeres Le Chatels, France
}

\and 
\author{F. Soubiran}
\affil{LGLTPE, 
Ecole normale sup\'erieure de Lyon, 69364 Lyon Cedex 07,  France\\
}

\date{}

\begin{abstract}
We present a new equation of state (EOS) for dense hydrogen/helium mixtures which covers a range of densities from
$10^{-8}$ to $10^{6}\,\gcc$, pressures from $10^{-9}$ to $10^{13}$ GPa and temperatures from $10^{2}$ to $10^{8}$ K. 
The calculations combine the EOS of Saumon, Chabrier \& vanHorn (1995) in the low density, low temperature molecular/atomic
domain, the EOS of Chabrier \& Potekhin (1998) in the high-density, high-temperature fully ionized domain,
the limits of which differ for H and He, and ab initio quantum molecular dynamics (QMD) calculations in the intermediate density and temperature regime, characteristic of pressure dissociation and ionization. The EOS for the H/He mixture is based on the so-called additive volume law and thus does not take
into account the interactions between the two species. A major improvement of the present calculations
 over existing ones is that we calculate the entropy over the entire density-temperature domain, a necessary quantity for stellar or planetary evolution calculations. The EOS results are compared with existing experimental data, namely Hugoniot shock experiments for pure H and He, and with
first principle numerical simulations for both the single elements and the mixture. 
This new EOS covers a wide range of physical and astrophysical conditions, from 
jovian planets to solar-type stars, and recovers the existing relativistic EOS at very high densities,
in the domains of white dwarfs and neutron stars.
All the tables are made publicly available.
\end{abstract}



\section{Introduction} 
\label{intro}

Understanding the thermodynamic properties of hydrogen (H) and helium (He) at high density, characterized by their equation of state (EOS),
is at the heart of numerous physical and astrophysical problems.
From the  point of view of fundamental physics, understanding the metalization of hydrogen has remained a major challenge since the pioneering work of Wigner \& Huntington (1935), more than 80 years ago.
The quest for its observational evidence remains so far unachieved but is in reach with both static and dynamic high pressure experiments, thanks,
for these latter, to the achievement of modern techniques such as Z pinch magnetically driven shock experiments (Knudson et al. 2004), spherically converging shock wave experiments (Belov et al. 2002, Boriskov et al. 2003) and intense laser driven planar shock wave experiments (Collins et al. 1998,
Hicks et al. 2009, Sano et al. 2011, Loubeyre et al. 2012, Brygoo et al. 2015). These experiments have revealed the principal Hugoniot of dense deuterium up to 200 GPa. 
Knowledge of the hydrogen and helium EOS is also central for inertial confinement fusion (ICF) and of course for the characterization of the interior or outer mechanical and
thermal structures of dense astrophysical bodies. These latter include low-mass stars (generically stars smaller than the Sun, for which the perfect gas EOS or the Debye-H\"uckel expansion is no
longer valid), brown dwarfs (objects not massive enough to sustain or even ignite hydrogen fusion in their core, whose mass distribution extends from about 0.07 $\msol$ down to a few Jupiter masses), giant
(solar and extrasolar) planets, but also the envelope of white dwarfs and the outer envelope and atmosphere of neutron stars.

In the meantime, ab initio numerical calculations of the properties of dense H and He, based either on quantum molecular dynamics (QMD), which combines molecular dynamics (MD) for
the heavy classical particles and density functional theory (DFT) to treat the quantum electrons, or Path Integral Monte Carlo (PIMC) or Molecular Dynamics (PIMD) can now be performed in the density-temperature domain
of interest (e.g. Militzer \& Ceperley 2000, Holtz et al. 2008, Militzer 2009, 2013, Militzer et al. 2001, Morales et al. 2010ab, Lorenzen et al. 2009, 2011,  Becker et al. 2014, Mazzolla et al. 2018, Sch\"ottler \& Redmer 2018), thanks to the enormous improvement in computer capacities. The widely used semi-analytical H/He model of Saumon \& Chabrier (Chabrier 1990, Saumon \& Chabrier 1991, 1992, Saumon et al. 1995 (SCvH))
can thus now be replaced by these calculations in the crucial domain of pressure ionization. Such an approach, combining ab initio calculations with the SCvH EOS in the low (mainly
atomic/molecular) and high (fully ionized) domains, has been used by various authors (Caillabet et al. 2011, Militzer \& Hubbard 2013, Becker et al. 2014, Miguel et al. 2016). These calculations, however, remain so far limited
in two aspects. Either they cover only a limited density-temperature range, precluding the use of an EOS over a significant physical or astrophysical domain, or they do not
provide the entropy. Indeed, while the pressure and internal energy are directly accessible to QMD or PIMC/PIMD calculations, the entropy is a much more cumbersome task, requiring a so-called thermodynamic
integration over a large number of temperature and density points. The knowledge of the entropy, however, is central in stellar evolution calculations (as the cooling history of a star directly derives from
the first principle of thermodynamics, $Q=dS/dt$) and even to determine the thermal structure of 
dense astrophysical
bodies since their interior is quasi isentropic, due to the onset of convection to carry out their 
internal heat flux\footnote{It must be kept in mind that only for adiabatic {\it reversible} process, such e.g. as convection, is an adiabat ($dQ$=0) equivalent to an isentrope ($dS=0$).}Ê. The thermal profile and the contraction rate, thus the
evolution of low-mass stars, brown dwarfs
and giant planets is indeed entirely determined by their entropy profile
(Chabrier \& Baraffe 2000). 

In the present paper, we follow the same method as mentioned above, by combining QMD calculations for pure hydrogen and helium with the SCvH (1995) 
and the Chabrier-Potekhin (1998) EOSs. As just mentioned, a striking advantage of the 
present calculations is that they provide the entropy over a wide temperature-pressure-density range, namely
$10^{-8}$ to $10^{6}\,\gcc$, $10^{-9}$ to $10^{13}$ GPa and $10^{2}$ to $10^{8}$ K, covering essentially the domain of all dense astrophysical bodies. 
The paper is organized as follows. Sections 2 and 3 describe the H and He EOS, respectively, and make comparison with available experimental data or ab initio calculations. The calculations for the H/He mixture
are described in \S4, examples of the tables are presented in \S5 while \S6 is devoted to the conclusion.

\section{The Hydrogen equation of state}
\label{EOSH}
\subsection{Construction of the EOS model}
\label{Hmodel}

Following the same procedure as Becker et al. (2014), our hydrogen EOS combines different calculations.
For $T\ge 1.1\times 10^5$ K and/or $\rho > 10\gcc$, the hydrogen becomes fully ionized and we use the Chabrier \& Potekhin (1998, CP98) EOS model,
based on the linear response theory to treat ion-electrons interactions. The CP98 EOS extends to very high temperatures or densities, when
electrons become relativistic and recovers the Potekhin \& Chabrier (2000) model EOS which handles the solid phase. The relativistic domain concerns essentially
neutron stars or white dwarfs.
For  $T< 1.1\times 10^5$ K, the EOS is divided into 3 density regimes, where we use 3 different EOS
calculations:
\begin{itemize}
\item $\rho \le 0.05\,\gcc$: SCvH EOS
\item $0.3< \rho\le 5.0\,\gcc$: EOS of Caillabet et al. (2011) 
\item $\rho> 10.0\,\gcc$: CP98 EOS
\end{itemize}
Between these limits, a bicubic spline interpolation is performed which ensures continuity of the functions and their two first derivatives.

\begin{figure}
\includegraphics[width=\linewidth]{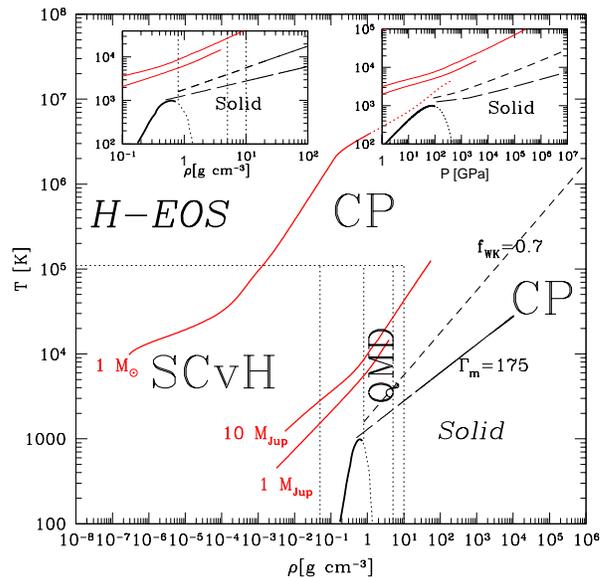}
\vspace{-2cm}
\caption{Temperature-density domain of the present EOS for hydrogen. The dotted
lines illustrate the $T$-$\rho$ domains corresponding to the different models or calculations combined to produce the final EOS (see text). Between these domains, bicubic spline interpolations have been used on the various thermodynamic quantities. The melting lines for H$_2$ (eq. (\ref{Hmelt})) and
H$^+$ (eq. (\ref{OCP})) are delimited by the thick and thin solid lines, respectively, in the lower right
corner (note that the line for H$^+$ is extrapolated beyond the validity of the OCP model for illustrative purposes only). The short-dashed line $f_{WK}=0.7$ corresponds to the limit of validity of the present calculations, due to ion
quantum effects (eq. (\ref{fwk})). The two insets focus on the liquid to solid and ion classical
to quantum
locations of the phase diagram. The EOS must not be used beyond these limits. 
Interior profiles for the Sun (1 $\msol$) and 1 and 10 $\mjup$ planets
at 5 Gyr (from Baraffe et al. 2003, 2015) are displayed in the figure to illustrate the domain of astrophysical applications. 
}
\label{fig1}
\end{figure}

Whereas, as mentioned in \S\ref{intro}, the calculations are performed over a vast density-temperature domain, namely $10^{-8}\le\rho\le 10^6\,\gcc$ and $10^2\le T\le10^8$ K, several limits are identified. 

(1) For $T\la 10^3$ K, hydrogen becomes solid over some pressure/density range. The melting line for
H$_2$ has been determined experimentally up to $T\simeq 1000$ K, $P\simeq 100$ GPa (Datchi et al. 2000, Deemyad \& Silvera 2008, Eremets \& Trojan 2009)
 and has been extrapolated up to about 300 GPa 
by the following fonctional form (Kechin 1995), based on QMD simulations (Bonev et al. 2004, Morales et al. 2010a, Caillabet et al. 2011\footnote{Note the typo in eqn. (41) of Caillabet et al. (2011), corrected in eqn.(\ref{Hmelt}) here.}):

\begin{eqnarray}
T_m=T_0(1+P/a)^{b}\,exp(-cP),
\label{Hmelt}
\end{eqnarray}
with $T_0=4.853$ K, $a=0.023$ GPa, $b=0.748$ and $c=0.0098$ GPa$^{-1}$ for H$_2$.
Note, however, that the turning point at $P\ga 100$ Mbar predicted by this form has not been confirmed unambiguously yet by experiments. The experimentally
determined H$_2$ melting
curve is identified in Fig. \ref{fig1} by the thick solid line and is continued by the
dotted line at higher density/pressure according to the above functional form. 
At higher temperatures and pressures, hydrogen becomes fully dissociated and
ionized, reaching the limit of the one component plasma model (OCP) whose melting line corresponds to $\Gamma_m=175$ (Potekhin \& Chabrier 2000), where $\Gamma=(Ze)^2/(a\kb T)=2.25\times 10^5(Z^2/A)(\rho^{1/3}/T)$ is the usual plasma coupling parameter. This yields the following melting line:

\begin{eqnarray}
\log \,T_m\approx 3.1+\frac{1}{3}\log\, \rho+2\, \log\, Z-\frac{1}{3} \log \,A,
\label{OCP}
\end{eqnarray}
as identified in Fig.  \ref{fig1}. Interestingly enough, extrapolating this line to lower
temperatures and densities (long-dash) nicely joins the H$_2$ melting line. Note, however, that this
line is just indicated for illustrative purposes and can not be considered as a rigorously determined melting line at high pressure. Indeed, at very
high densities/pressures, quantum diffraction effects between protons become significant
and the classical OCP model becomes invalid (see below). 

(2) At low temperature and high density, quantum (diffraction) effects between ions become important.
In the CP98 model, these effects are treated within the $\hbar^2$ Wigner-Kirkwood expansion to second order. This yields the free energy quantum correction
$f_{WK}=F_{WK}/N\kb T=\eta^2/24$,
 where $\eta=\hbar \Omega_p/\kb T\approx 7.71\times 10^{3} \rho^{1/2}T^{-1} Z A^{-5/3}$ and $\Omega_p=(4\pi(Ze)^2n_i/M_i)^{1/2}$ denotes the ion plasma frequency. 
 For $f_{WK}\ga 0.7$, the CP98 model has been found to
 become of dubious validity and then the present EOS can not be used beyond
this limit. This limit corresponds to 
\begin{eqnarray}
	\log\, T\approx 3.3+\frac{1}{2}\log\, \rho+ \log\, Z-\frac{5}{3} \log \,A,
\label{fwk}
\end{eqnarray}
 indicated by the short-dash line in Fig.  \ref{fig1}. 
Beyond this limit, the treatment of
quantum effects requires fully quantum numerical calculations such as PIMC or PIMD. Such a quantum domain for hydrogen, however, does not concern any astrophysical body
(see, e.g. Chabrier 1993).

The hydrogen QMD calculations in the intermediate density regime (see above) are based on Caillabet et al. (2011), 
which gather QMD simulations by Holst et al. (2008), coupled electron-ion Monte Carlo (CEIMC) calculations by Morales et al. (2010a),
and PIMC calculations by Militzer \& Ceperley (2000). These calculations have been supplemented by further QMD calculations for our present purpose.
The excess free energy, $F_{ex}$, was fitted by a functional form similar to the one proposed in Chabrier \& Potekhin (1998), which accurately recovers all appropriate limits. The accuracy of this analytical parameterization was verified by the fact that its
temperature derivative properly recovers the excess internal energy, $U_{ex}$, obtained in the simulations.
In the present calculations, however, we found out that, whereas the fit for $\fex$ used in Caillabet
et al. (2011) correctly recovers the H$_2$
melting curve, it becomes less accurate away from these conditions. Therefore, in the present
calculations, we have modified the $d(\rho)$ parameter of the fit given in eqn.(24) of
Caillabet et al. (2011) in various density domains in order to recover the ab initio calculations of Morales et al. (2010a). The results will be illustrated in \S\ref{calc} below for H and in \S4 for the H/He mixture.

As mentioned above, the EOS is calculated initially in a $T$-$\rho$ domain, appropriate to QMD or PIMC calculations,
 and then transformed into a $T$-$P$ one by bicubic interpolation procedures. In \S\ref{calc}, we will make extensive comparisons between our results and available numerical results from ab-initio simulations for several
thermodynamic quantities in order to verify the validity of these EOS calculations.

\subsection{Comparison with experimental results}
\label{Hugo}

The validity of the EOS of hydrogen, or its isotope deuterium, can be first assessed by comparison with high-pressure Hugoniot experiments.
As mentioned in the introduction, these latter include different techniques. The original discrepancies
between these various data sets have been significantly reduced when using a revised EOS of quartz for the impedance-matching in the case of laser-compression experiments (Knudson \& Desjarlais 2009), and all
results now agree reasonably well to provide a robust compression Hugoniot curve up to about 200 GPa. The precision of these measurements
has been improved recently with magnetically accelerated flyer plate experiments on deuterium, reaching a precision of $\sim$1.5\%-$1.9\%$ in density along the Hugoniot, and carrying out reshock measurements from these Hugoniot states that provide off-Hugoniot data in the $\sim100$-200 GPa and $\sim5000$-15000 K regime (Knudson \& Desjarlais 2017).
Furthermore, experiments combining
static and dynamic methods, generating laser-driven planar shock waves in precompressed samples of different initial densities have
allowed to explore a larger domain off the principal Hugoniot, probing the EOS of hydrogen isotopes over an even larger pressure-temperature domain, directly probing the conditions in giant planet interiors (Loubeyre et al. 2012, Brygoo et al. 2015).

In order to compare our EOS calculations with experiments, 
we have calculated Hugoniot pressure-density and pressure-temperature curves for D$_2$ and H$_2$. Postshock conditions are calculated from mass, momentum and energy conservation across the shock by finding solutions from the tabulated
EOS that satisfy the Hugoniot relation (Zel'dovich \& Raizer 2002) for given initial conditions ($\rho_0, P_0, {\tilde E}_0$), where these quantities denote respectively the mass density,
pressure and specific internal energy\footnote{Note that for the shock velocities
under consideration, $U_S\approx 10$-50 km s$^{-1}$, radiative effects in the energy balance are negligible.}:

\begin{eqnarray}
({\tilde E}-{\tilde E}_0)+ \frac{1}{2} (P+P_0)\times(\rho^{-1}-\rho_0^{-1})=0
\label{Hug}
\end{eqnarray}

\begin{figure}
\includegraphics[width=\linewidth]{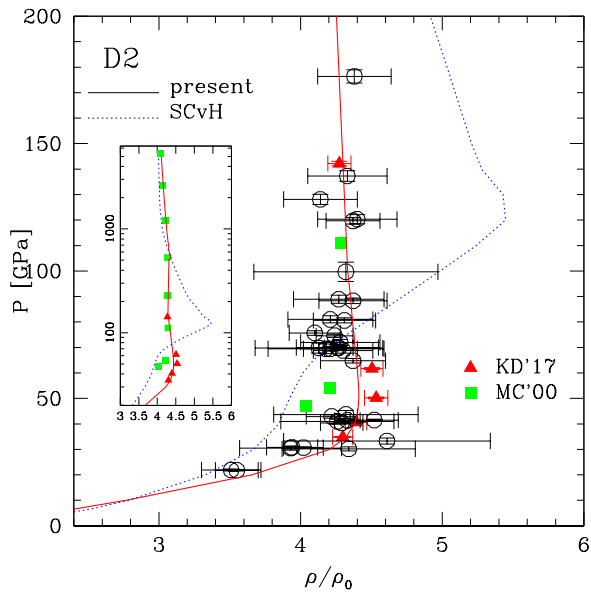}
\vspace{-2cm}
\caption{Shock pressure vs density along the deuterium Hugoniot curve. Solid triangles: results by Knudson \& Desjarlais (2017) for initial temperature
and density $T_0=20$ K and $\rho_0=0.167\gcc$, respectively. Empty circles: reanalyzed shock data obtained from various experiments (see text) rescaled to the
same initial density (data from Knudson \& Desjarlais (2017)). Solid squares: PIMC calculations of Militzer \& Ceperley (2000). Solid line: present calculations; dotted line: SCvH EOS.}
\label{fig2}
\end{figure}

Figure \ref{fig2} compares the Hugoniot compression curve for deuterium obtained with our EOS with the most recent data of Knudson \& Desjarlais (2017), which include also some of the aforementioned experimental results
data for an initial state $\rho_0(D_2)=\,0.167\gcc$ at $T_0=20$ K (the results originally obtained for a slightly larger initial density have been rescaled
accordingly (see Knudson \& Desjarlais (2017) for details)). To calculate the deuterium Hugoniot curve, we have rescaled the hydrogen EOS by a factor $2$ in density, but proper quantum corrections on the energy are taken into account.
Since D$_2$ and H$_2$ have similar molar volumes at these conditions, the Hugoniot curves are nearly identical for these two elements. 
Our EOS is in excellent agreement with the data, as already noted in Caillabet et al. (2011),
including with the most recent experiments.
The maximum discrepancy occurs at $P=50$ GPa
and amounts to $\sim 3 \%$ on the density.
In contrast, the SCvH EOS is less compressible in the low pressure domain and more compressible at higher pressures. Since the compression peak corresponds to the domain of molecular dissociation
(energy goes into the breaking of internal levels and molecular bonds, yielding an increase of $\rho/\rho_0$), 
this behaviour reflects a well known shortcoming of this EOS, which underestimates H$_2$ pressure dissociation. This stems essentially from the too stiff H-H and H$_2$-H potentials used in the Saumon-Chabrier theory, which do
not include the softening due to N-body interactions, in contrast to the case of the
 H$_2$-H$_2$ potential (see Saumon \& Chabrier 1991). Indeed, high-pressure experiments at this time were not reaching high enough
pressures to explore the dissociated regime, and thus could not provide experimental 
guidance to derive softened potentials for interacting atomic species.
The inset in Fig. \ref{fig2} displays the comparison between the present calculations and the PIMC simulations
by Militzer \& Ceperley (2000) at higher pressures.

\begin{figure}
\includegraphics[width=\linewidth]{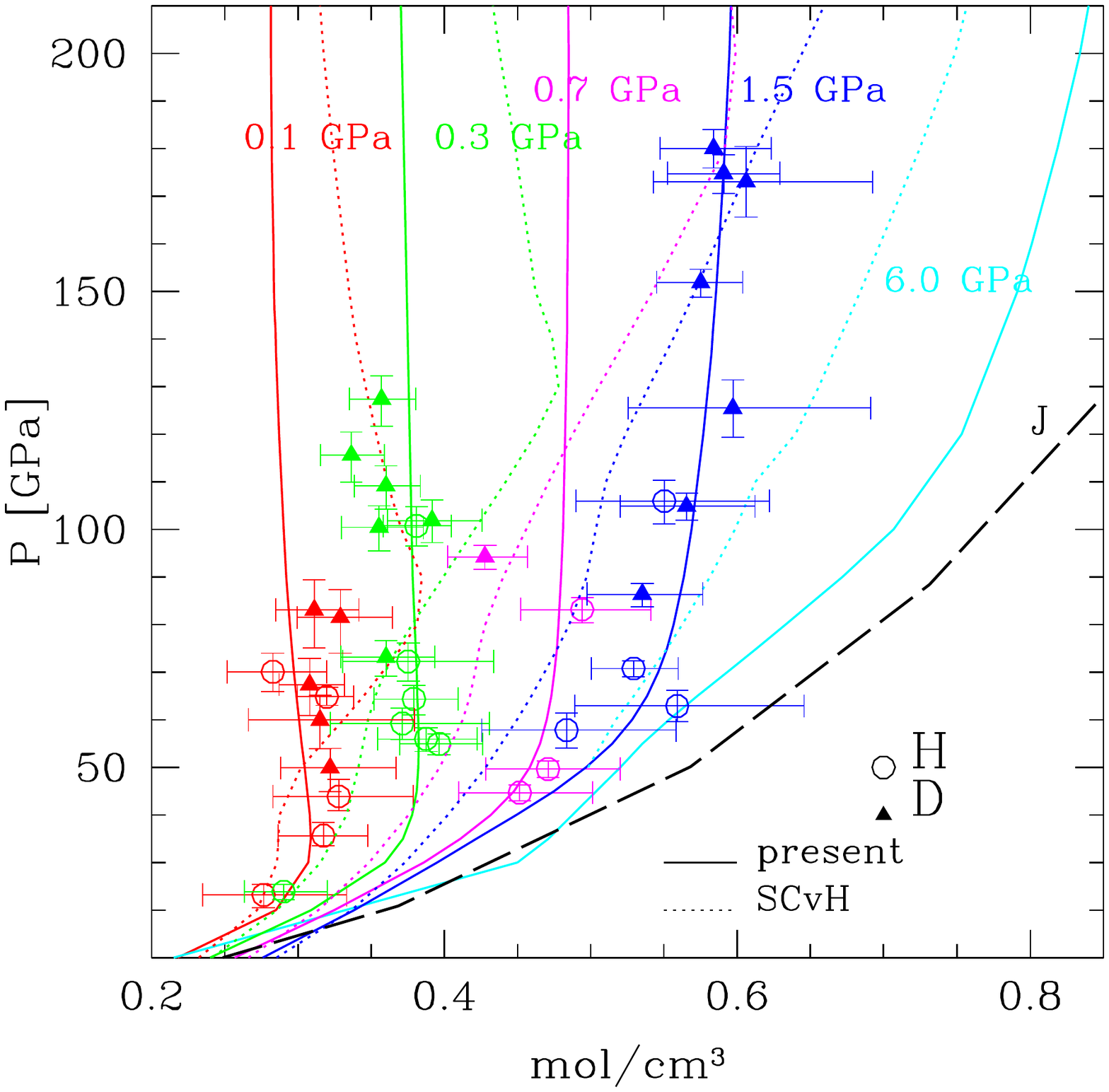}
\vspace{-2cm}
\caption{Hydrogen (empty circles) and deuterium (solid triangles) shock pressure vs density along the Hugoniot for $T_0=297$ K and 
various precompressed
initial conditions, namely 0.1, 0.3, 0.7, 1.5 and 6.0 GPa, as labeled in the figure. Data: Brygoo et al. (2015); solid line: present calculations; dotted line: SCvH EOS.
A Jupiter internal isentropic profile (for $x_{He}\simeq 0.08$), is portrayed by the long-dashed line.}
\label{fig3}
\end{figure}

Figure \ref{fig3} compares our theoretical Hugoniots for H$_2$ and D$_2$ 
with the ones obtained for various pre-compressed initial states
(Loubeyre et al. 2012, Brygoo et al. 2015). Initial states have pressures $P_0=0.16$ GPa, 0.3 GPa, 0.7 GPa and 1.5 GPa at 297 K, respectively. Again, the agreement between the data and the present EOS is excellent for all series of experiments. Also shown for comparion is a predictive Hugoniot
calculated for an initial pressure $P_0=$6.0 GPa, as planed with future high-pressure experiments, as well as a typical Jupiter internal isentrope for a helium number fraction $x_{He}\simeq 0.08$ (mass fraction $Y\simeq 0.25$ (see \S4)). As seen in the figure, conditions along this Hugoniot are very close to or intercept Jupiter's internal density and temperature profiles (assuming an isentropic thermal structure), respectively, 
notably  in the crucial $\sim$Mbar pressure ionization region, and thus directly probe Jupiter's deep interior.
 The same aforementioned general behaviour of the SCvH EOS, i.e. 
underestimated molecular hydrogen pressure dissociation,
 is observed for all Hugoniots
and is particlarly striking along the 6.0 GPa one.

\begin{figure}
\includegraphics[width=\linewidth]{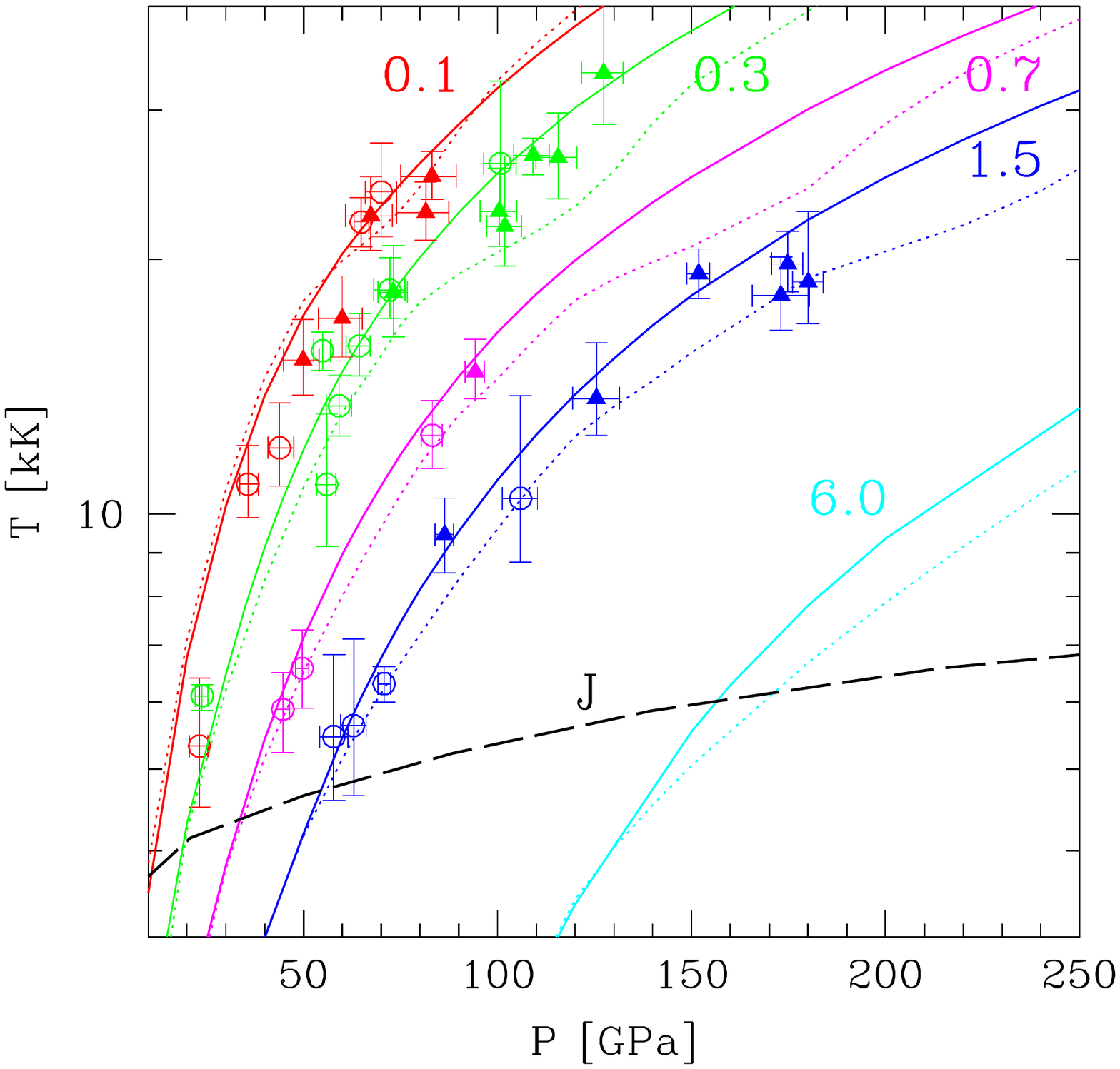}
\vspace{-2cm}
\caption{Hydrogen (empty circles) and deuterium (solid triangles) shock temperature vs pressure along the Hugoniot for the same precompressed
initial conditions as in Fig. \ref{fig3}, namely 0.1, 0.3, 0.7, 1.5 and 6.0 GPa from top to bottom. Same labeling as in Fig. \ref{fig3}.}
\label{fig4}
\end{figure}

Figure \ref{fig4} portrays the temperature-pressure curves along the Hugoniots for the same sets of experiments. Interestingly enough, the difference between
the present EOS, which includes ab initio simulations, and the semi-analytic SCvH EOS is much smaller than for the $P$-$\rho$ compression curves. We note, however, that molecular
dissociation in the SCvH EOS not only occurs at too high pressures, as mentioned above, but
takes place very abruptly, as shown by the kinks in the dotted curves, yielding cooler temperatures at given pressure as energy goes into molecular
dissociation instead of raising $\kb T$.
Experiments (Loubeyre et al. 2012), in contrast, have revealed that reflectivity, then electrical conduction, increases gradually along the Hugoniot above about $\sim$ 5000 K, before reaching a plateau,  
reflecting the dissociation and ionisation of molecular hydrogen H$_2$ and suggesting that
this process, under the conditions probed by present Hugoniot experiments, occurs continuously. This is in agreement with 
ab initio calculations, which show an increasing conductivity along the Hugoniot, but predict
a discontinuous molecular-ionic transition around P$\simeq 100$ GPa at cooler temperatures, in the range $T\simeq 2000$-6000 K (Morales et al. 2010b, Mazzola et al. 2018). 

\begin{figure}
\includegraphics[width=\linewidth]{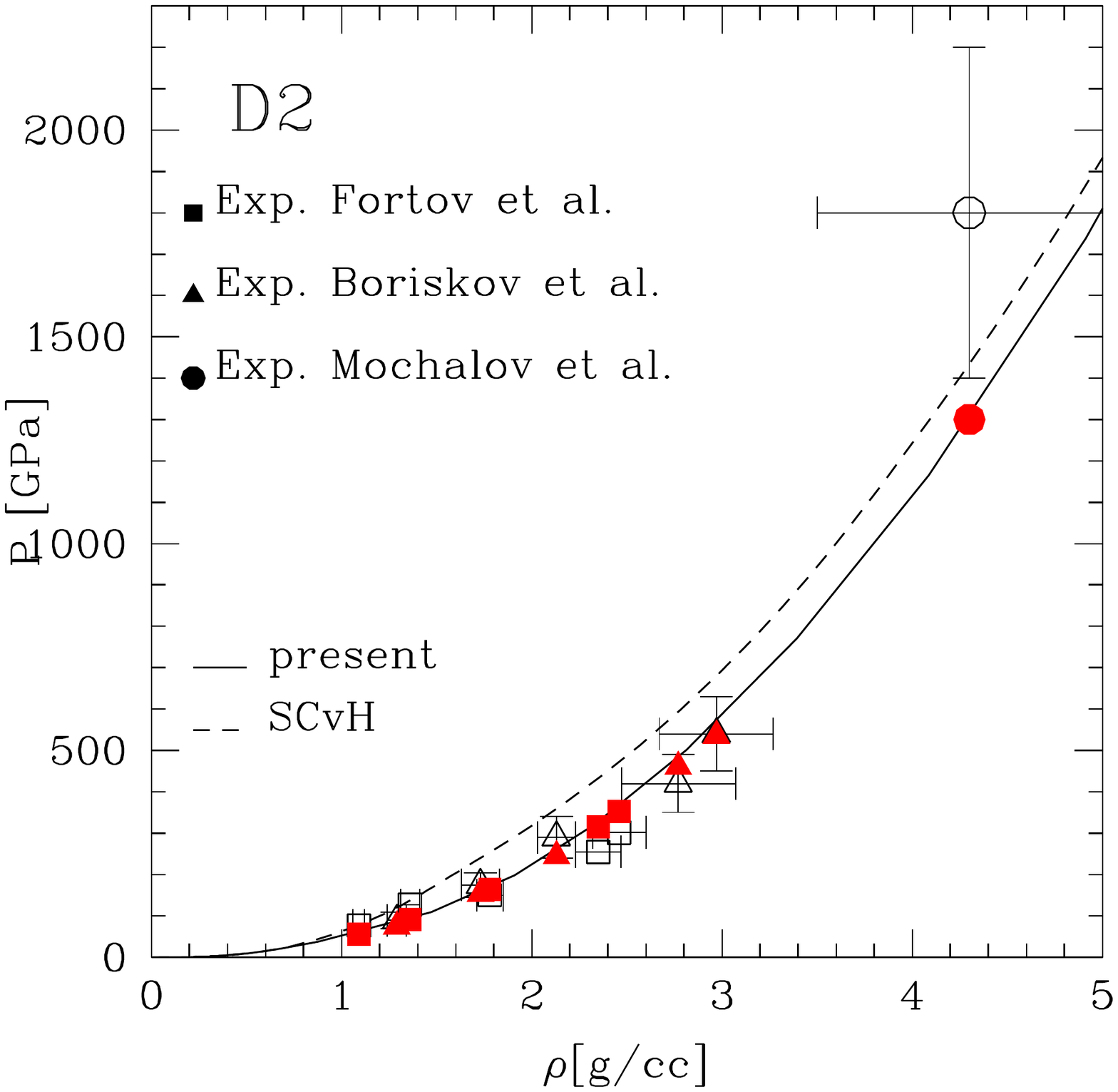}
\vspace{-2cm}
\caption{Isentropic compression of deuterium, for initial temperature
and density $T_0=283$ K and $\rho_0=0.104\gcc$, as in Fortov et al. (2007). 
Squares: Fortov et al. (2007); triangles: Boriskov et al. (2011); circles: Mochalov et al. (2010). Empty symbols: pressures determined in the experiments with their model EOS;
solid symbols: pressures obtained by Becker et al. (2013). Solid curve: present EOS; dashed curve: SCvH EOS.}
\label{fig5}
\end{figure}

Another important experimental constraint on the EOS comes from quasi-{\it isentropic} ramp compression of hydrogen or deuterium. Those experiments have direct astrophysical applications since,
as will be discussed later, the interiors of low-mass objects are entirely convective such that their internal temperature profile follows an isentrope. Dynamic quasi-isentropic shock wave experiments using high explosives on deuterium have been carried out up to about 1500 GPa (15 Mbar) and densities about 4.5 $\gcc$, directly probing jovian planet deep interiors. While, in some cases, both the density $\rho$ and the pressure $P(\rho)$ were measured simultaneously (Boriskov et al. 2011), in other experiments only the densities were measured while the pressure was determined afterwards from a hydrocode with a model EOS (Fortov et al. 2007). An extension of these latter experiments was
carried out by Mochalov et al. (2010) up $\rho=108\times \rho_0=4.3\gcc$, reaching an unprecedented experimental pressure for D$_2$ of 1800 GPa (18 Mbar). In all cases, the temperatures were determined from a model EOS. As noted by Becker et al. (2013), however, the experimental points of Fortov et al. and Mochalov et al. were found not to lie on the same isentrope,
questioning the validity of the results, at least of the model-dependent pressure determination from the measured density. Following Becker et al. (2013), we have calculated the
isentropic compression path obtained with our and SCvH EOS's, respectively, starting from Fortov et al. (2013) model-independent initial condition for D$_2$, $\rho_0=0.04 \gcc$, $T_0=283$ K. According to the present and SCvH EOSs, this corresponds to an entropy $S=9.9\,\kb$/atom. 
The result is displayed in Fig. \ref{fig5}. The empty symbols are the quoted experimental $\rho$-$P$ determinations while the solid symbols
correspond to the values obtained by Becker et al. (2013, Table I) with their EOS for the above initial conditions. As seen in the figure, our isentrope agrees very well with the values
obtained by Becker et al., with then all the experimental results lying on the same isentrope. We see in particular that the pressure rises continuously with the density along the
isentrope, with no sign of discontinuity due to a first-order phase transition in this regime, as suggested by Fortov et al. (2007). Interestingly enough, we see that the SCvH EOS
predicts larger pressures for a given density (by about $\sim10\%$ at 2 $\gcc$), i.e. a significantly ($> 20\%$) warmer isentrope.

\subsection{Comparison with ab initio calculations}
\label{calc}

\begin{figure}
\includegraphics[width=\linewidth]{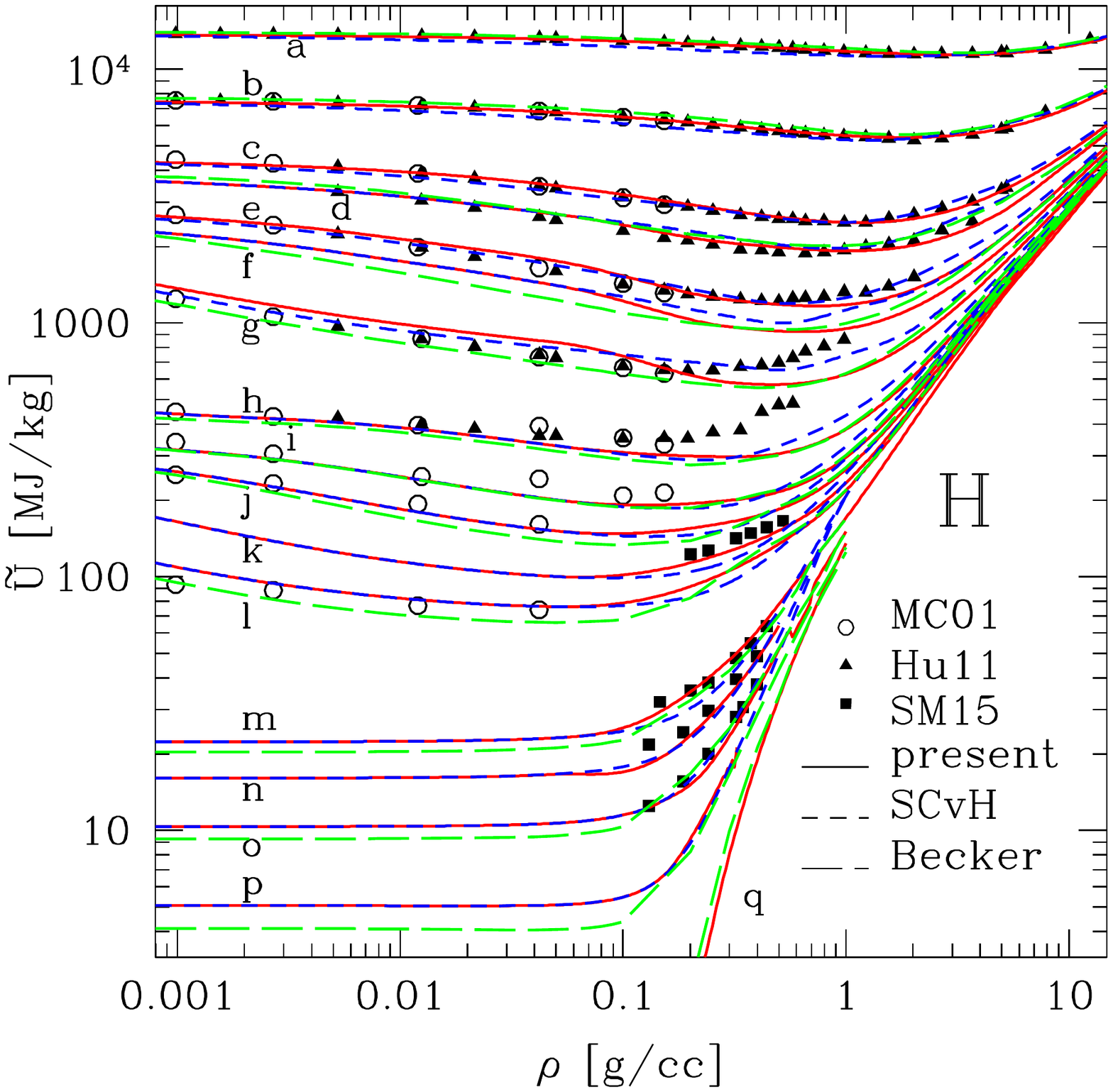}
\vspace{-2cm}
\caption{Specific internal energy vs density for several isotherms for hydrogen, labeled as follows: (a) 500 kK, (b) 250 kK, (c) 125 kK, (d) 100 kK, (e) 62 kK, (f) 50 kK, (g) 30 kK, (h) 15 kK, (i) 10 kK, (j) 8 kK, (k) 6 kK, (l) 5 kK, (m) 2 kK, (n) 1.5 kK, (o) 1 kK, (p) 500 K, (q) 100 K.
Empty circles: Militzer \& Ceperley (2001, MC01); solid triangles: Hu et al. (2011); solid square: Soubiran \& Militzer (2015, SM15).
(red) Solid lines: present calculations; (blue) short-dashed lines: SCvH; (green) long-dashed lines: Becker et al. (2014). }
\label{fig6}
\end{figure}

\begin{figure}
\includegraphics[width=\linewidth]{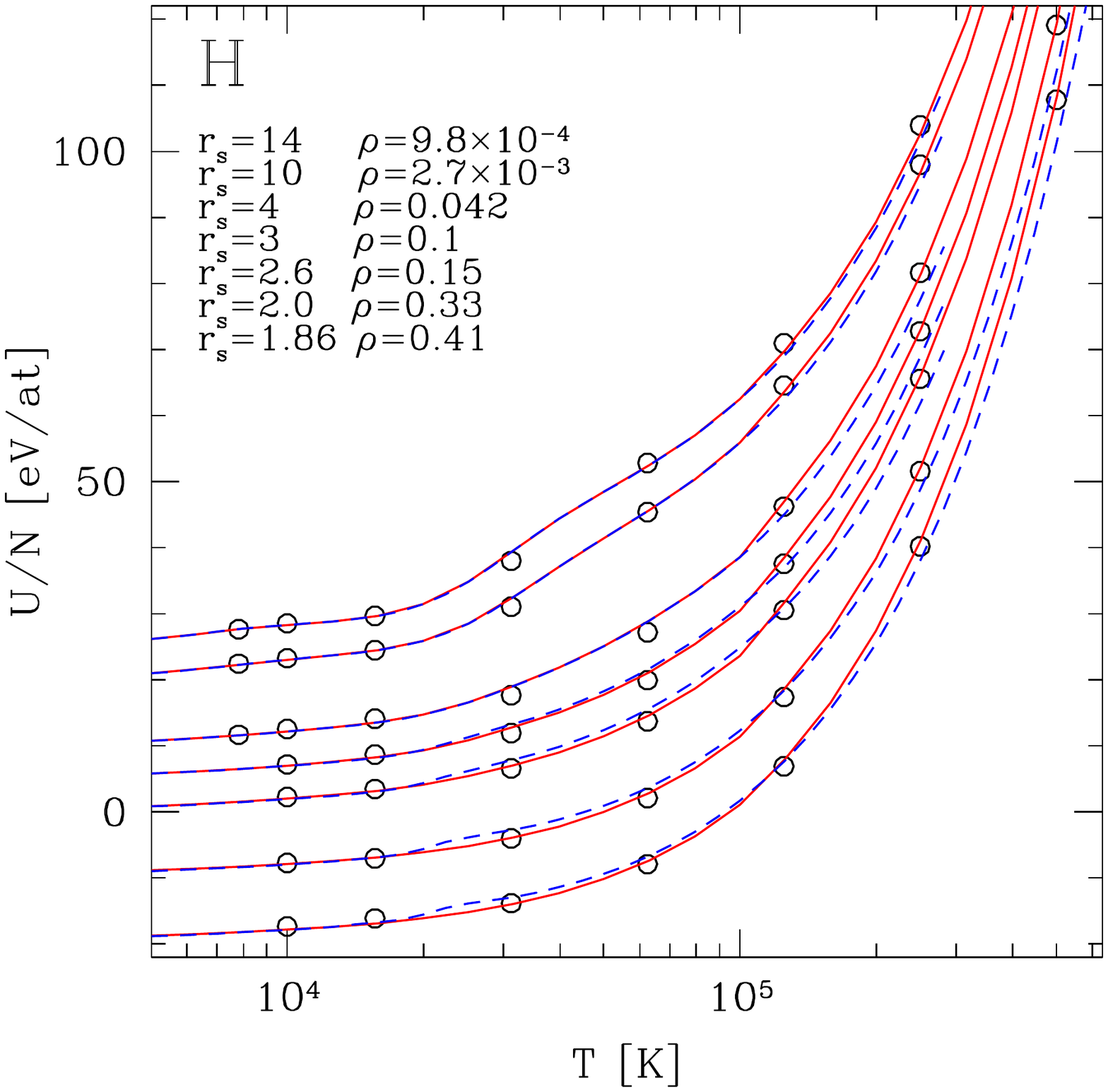}
\vspace{-2cm}
\caption{Hydrogen internal energy per atom vs temperature for several isochores (as labeled from top to bottom in the figure).
Empty circles: Militzer \& Ceperley (2000, 2001, MC).
Solid lines: present calculations; (blue) short-dashed lines: SCvH. The zero of energy is the same as MC but for sake of clarity, the curves have been shifted arbitrarily.}
\label{fig7}
\end{figure}

Figure \ref{fig6} compares the specific internal energy, $\ubar$, as a function of density for the present calculations with available PIMC (Militzer \& Ceperley 2001 (MC01), Hu et al. 2011)
and QMD (Soubiran \& Militzer 2015, SM) simulations over the available temperature-density ranges.
We also make comparisons with the QMD simulations of Becker et al. (2014) over a larger density
domain. Note that some of these simulations encompass the domain of hydrogen dissociation and ionization, i.e. $\rho \sim 0.5$-$5\gcc$, $T\sim 3000$-50000 K.
All results are rescaled to the zero of energy of the present EOS, which is the same as in SCvH, namely the ground state of the H$_2$ molecule. In all the domain explored by MC01 PIMC simulations, we note the good agreement between {\it all different calculations}, including SCvH and these simulations. 
Clearly, these simulations do not probe a density regime where differences between the various EOS due to the treatment of hydrogen dissociation and ionization arise.
The Hu et al. (2011) PIMC and Soubiran \& Militzer (2015) simulations, in contrast, reach higher densities and enter the crucial dissociation/ionization regime. The agreement between the present calculations and these simulations is excellent.
We notice, however, the surprising behaviour of the Hu et al. (2011) calculations at high density for the 15 kK, 30 kK and 62 kK isotherms. 
We need to stress here that, for these ($T,\rho$) conditions, the temperature is of the order of the electron Fermi temperature ($\theta=T/T_F\simeq 1$). Under such conditions, Monte Carlo samplings are known to be extremely inefficient and can lead to unreliable results.
We also notice an energy shift (i) between the PIMC and Becker et al. (2014) energy on one hand and the present and SCvH ones on the other hand for the 30000 K isotherm for $\rho \lesssim 0.1\gcc$ and (ii) between Becker et al. and the present or SCvH calculations for the
coolest isotherms even at very low densities. The shift at $T=30000$ K most likely stems from the underestimated 
H$_2$ dissociation in the SCvH EOS and thus in the present one below $\sim 0.3\gcc$, due to the interpolation
procedure (see \S2.1). Recall that our QMD calculations only extend down to 0.2 $\gcc$. The maximum discrepancy, however, is about 15\% around $\sim 0.05\gcc$ and becomes negligible below $\sim 0.01\gcc$. 
For this temperature, the PIMC simulations predict 57\%/43\%  H$^+$/H ionization fractions, with $x_{H_2}=0$ 
at $\rho=2.7\times 10^{-3}\gcc$ (see Table I of MC01) whereas SCvH predict 54\%/46\% H$^+$/H, with $x_{H_2}=0$, quite a good agreement.
The shift for the coolest isotherms,
notably at very low density, between Becker et al. (2014) and the SCvH and present calculations is more surprising as at these densities thermal dissociation and ionization, when they take place, are well described by the Saha equation,
a limit correctly recovered by SCvH. 
For $T\le 2000$ K, for which H$_2$ rotational levels are excited, but not vibrational ones, the SCvH correctly recovers
the perfect gas limit, ${\tilde U}=(5/2)\times\kb T/\mu=1.03\times 10^{-2}\times T$ MJ kg$^{-1}$, where $\mu=A_{H_2}\times \mh$, with $\mh=1.660\times 10^{-27}$ kg the atomic unit mass, which does not seem to be the case of the Becker et al. EOS.
Note in passing that the spin dependence of the H$_2$ molecule (ortho- and para-hydrogen) is correctly accounted for in the Saumon-Chabrier
theory (see Saumon \& Chabrier 1991).

All curves exhibit a sharp rise above $\sim 0.1$-5$\gcc$, depending on the temperature.
This corresponds to the increasing (repulsive) interactions between hydrogen molecules and/or atoms and then to
pressure ionization and the onset of the electron degeneracy contribution. Not surprisingly, then,
the most noticeable differences between the present, SCvH and Becker et al. (2014) EOSs occur in this domain, 
a regime covered by QMD simulations in both the present and Becker et al. EOSs but described by a semi-analytical model in SCvH.
Generally speaking, the SCvH EOS overestimates the internal energy compared with the simulations in this domain, except for the lowest temperatures where it first underestimates and then overestimates it.
This reflects the now well identified shortcoming of
the Saumon-Chabrier (SC) theory that overestimates the density domain of stability of molecular hydrogen, and then predicts a too abrupt ionization,
globally underestimating the domain of hydrogen dissociation/ionisation. 
This behaviour was already noted on the Hugoniot experiments (\S\ref{Hugo}), and arises essentially on one hand from 
the too stiff
interatomic potential compared to the intermolecular one in the SC theory, as mentioned previously, and also from the fact that atomic and molecular ionization, i.e. the Stark effect, are underestimated in the theory (see Saumon \& Chabrier 1991, 1992).

Figure \ref{fig7} displays similar comparisons for the energy per atom (as in MC01) as a function of temperature
between PIMC (Militzer \& Ceperley 2000, 2001) and the present and SCvH
EOSs along the available isochores. As mentioned above, within this density range, all types of calculations are in good agreement.
We note a slight departure between SCvH and the simulations for the highest density in the temperature range $\sim 30000$-80000 K, 
as already noted
by Militzer \& Ceperley (2001). This again reflects the imperfect treatment of ionization in the SC model. The deviations, however, remain modest, within at most $\sim 6\%$. In contrast, the present calculations agree very well with the MC01 PIMC results.

\begin{figure}
\includegraphics[width=\linewidth]{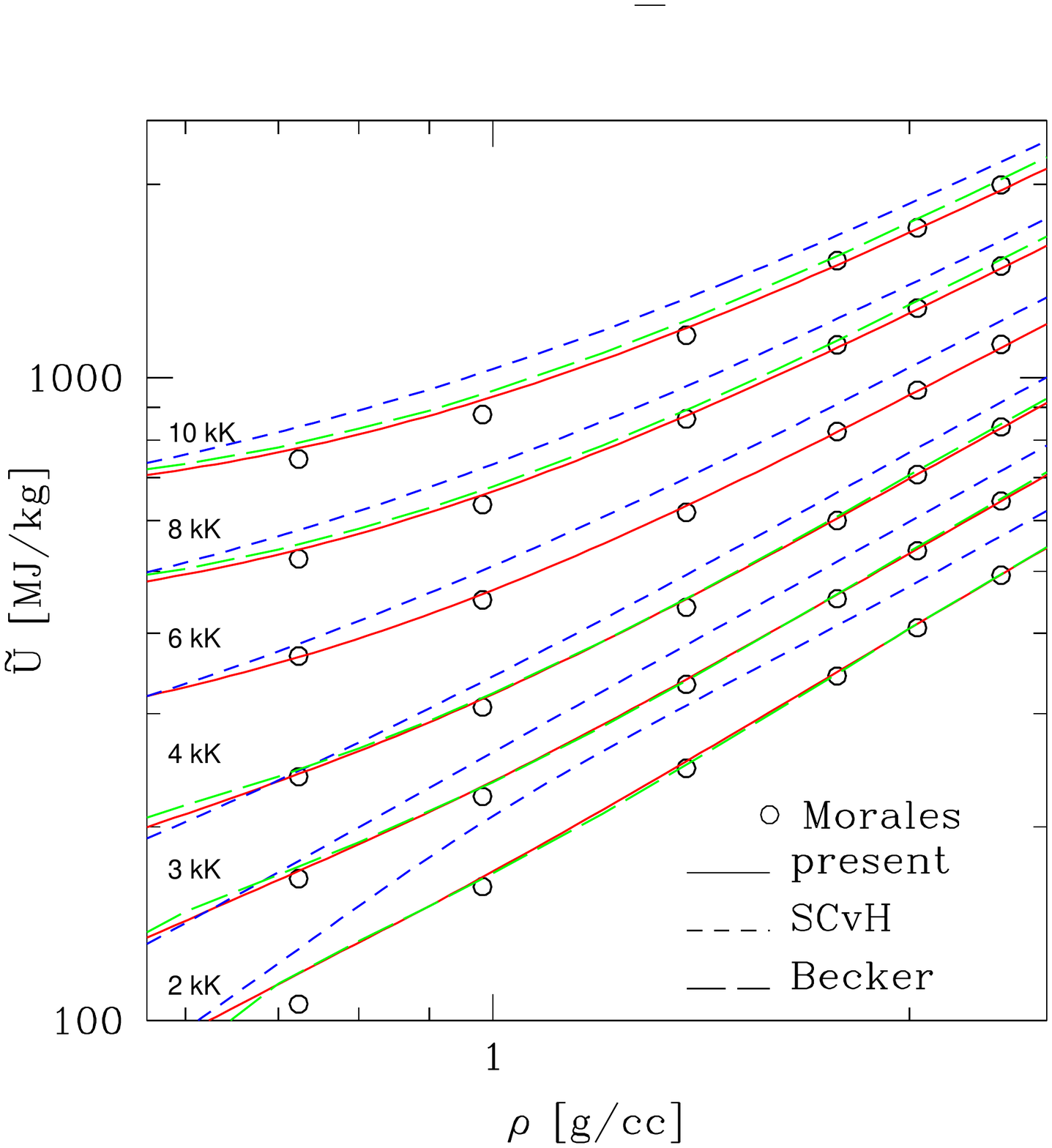}
\vspace{-2cm}
\caption{Specific internal energy vs density for hydrogen for several isotherms:
comparison with the CEIMC simulations of Morales et al. (2010a).
Solid lines: present calculations; (blue) short-dashed lines: SCvH; (green) long-dashed lines: Becker et al. (2014). 
For the sake of clarity all curves have been shifted upward by an arbitrary constant.}
\label{fig8}
\end{figure}

Figure \ref{fig8} portrays now similar comparisons with another set of first principle
simulations, namely the Coupled-Electron-Ion Monte-Carlo (CEIMC) calculations of 
Morales et al. (2010a), probing the very pressure-dissociation/ionization regime, between
$\rho=0.7\,\gcc$ and 2.4 $\gcc$. Remember that in this regime, both the present and Becker et al. (2014) EOS calculations rely on QMD
simulations. We first note the excellent agreement between QMD-based and
CEIMC results in the probed temperature-density range, which gives confidence that both types of 
methods can handle this crucial
density regime. We also note the strong departure between the SCvH EOS and these results, by
as much as 25\%, for the same reasons as mentioned previously.

\begin{figure}
\includegraphics[width=\linewidth]{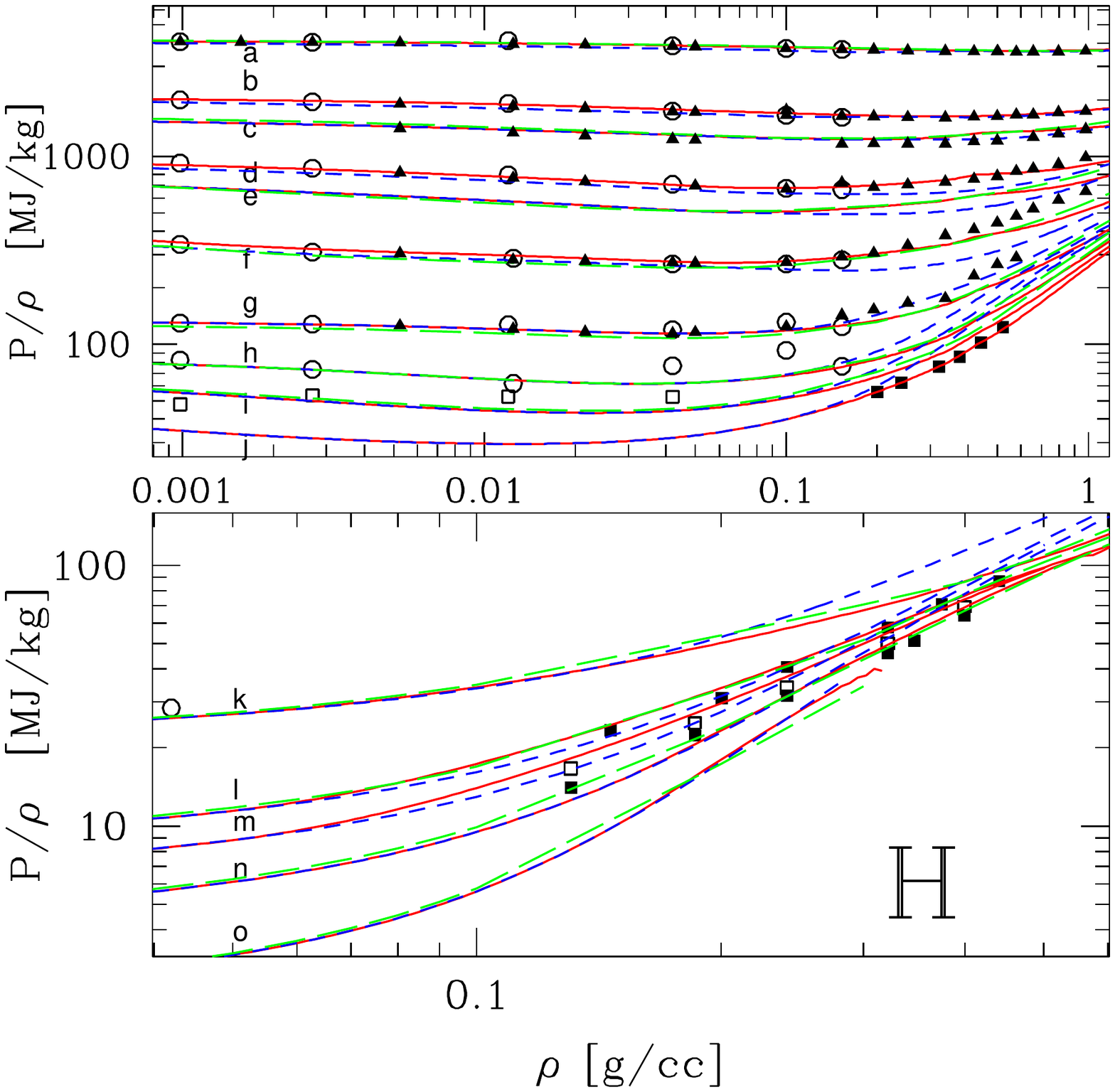}
\vspace{-2cm}
\caption{Comparison of pressure vs density for some of the same isotherms as in 
Fig. \ref{fig6}:  (a) 250 kK, (b) 125 kK, (c) 100 kK, (d) 62 kK, (e) 50 kK, (f) 30 kK, (g) 15 kK, (h) 10 kK, (i) 7.8 kK, (j) 6 kK, (k) 5 kK, (l) 2 kK, (m) 1.5 kK, (n) 1 kK, (o) 500 K. Same labeing as in Fig.\ref{fig6}. To avoid confusion with the nearest isotherms, the
MC01 data points for $T=7812$ K and the 
SM15 ones for $T=1.5$ kK are displayed with empty squares.}
\label{fig9}
\end{figure}

\begin{figure}
\includegraphics[width=\linewidth]{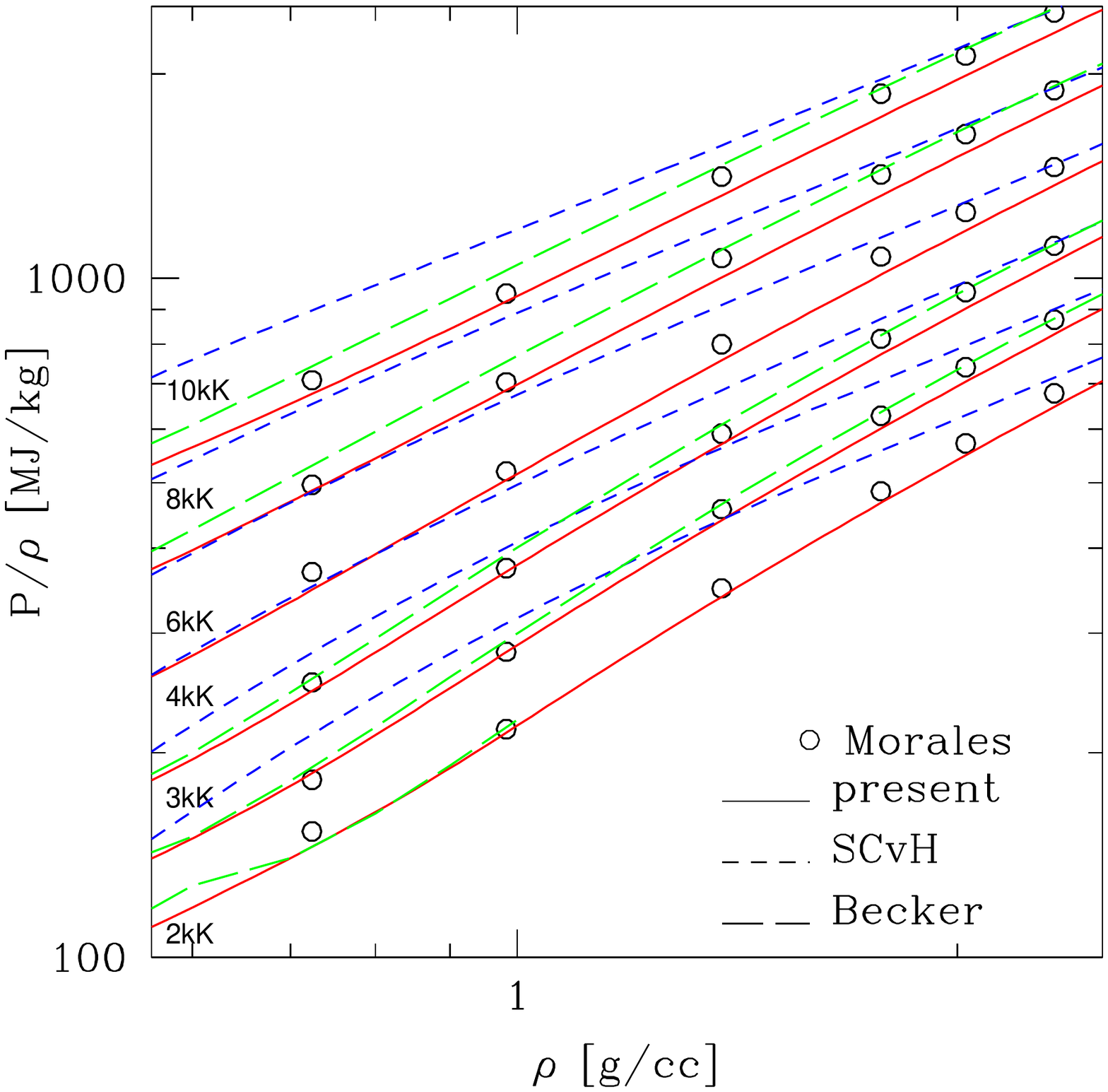}
\vspace{-2cm}
\caption{Same as Fig. \ref{fig8} for the pressure. 
As in Fig. \ref{fig8}, all curves have been shifted by a constant for clarity.}
\label{fig10}
\end{figure}

\begin{figure}
\includegraphics[width=\linewidth]{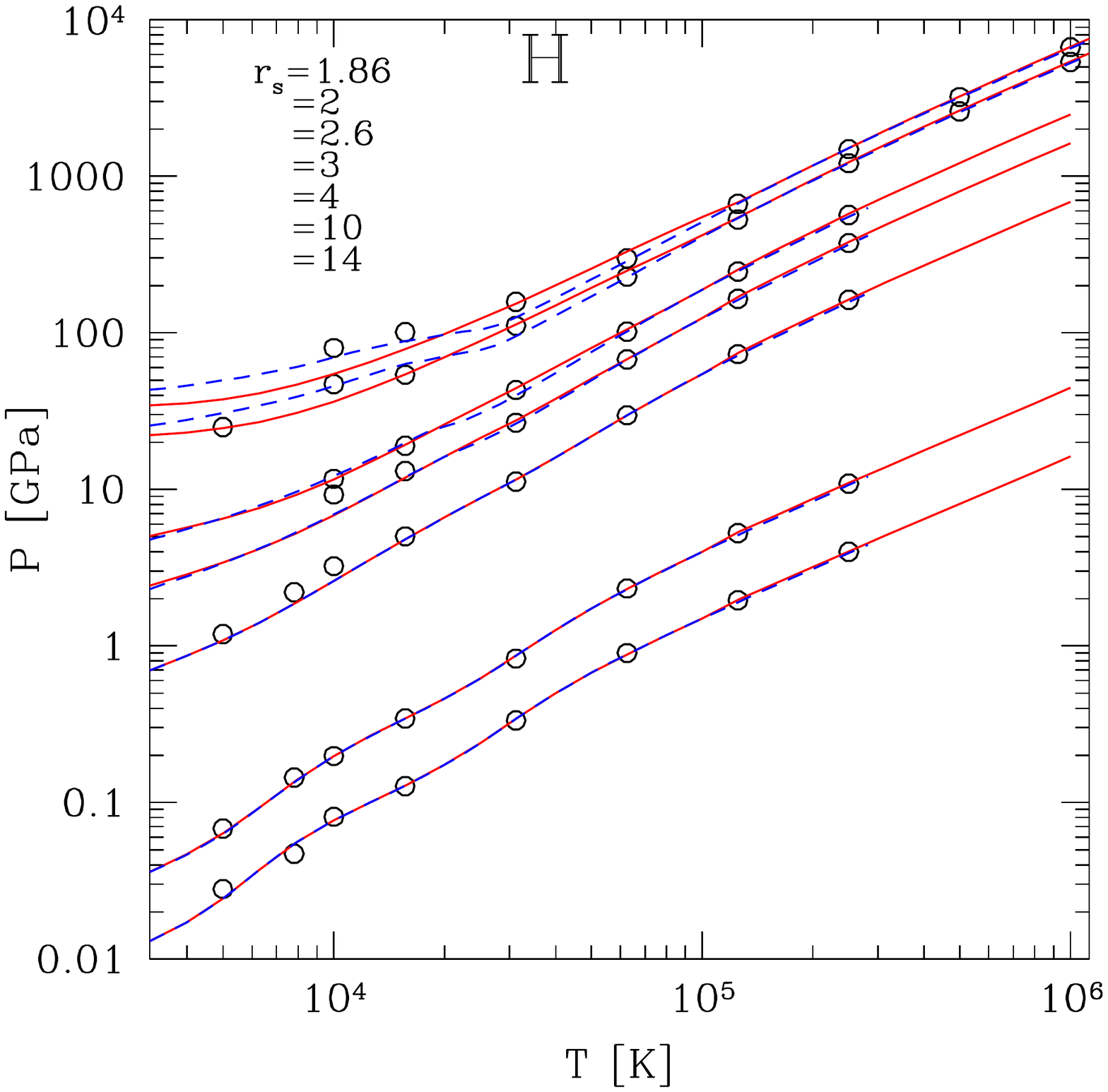}
\vspace{-2cm}
\caption{Same as Fig. \ref{fig7} for the
pressure. }
\label{fig11}
\end{figure}

Figures \ref{fig9} and \ref{fig10} compare the pressure, more precisely $P/\rho$ to highlight non-ideal contributions, as a function of density of the present EOS with the same set of 
simulations. 
As seen in Fig. \ref{fig9}, we first note that, as for the internal energy, all EOS calculations agree very well
with the simulations below about 0.1 $\gcc$. Above this value, the SCvH EOS significantly overestimates the pressure for $T\la 15$ kK and underestimates it at larger temperatures, by as much as 25\% around
1 $\gcc$, highlighting again the approximate
treatment of pressure dissociation and ionization in the
SC theory.
In contrast, the present EOS is in excellent agreement with
the Soubiran \& Militzer (2015) simulations, 
for the available isotherms, and with the Hu et al. (2011) ones for $T>60000$ K. For these latter, 
we note the same spurious behaviour for $T=15000$ K and 30000 K, as for the energy, which confirms the dubious validity of these results in the partially degenerate
 domain ($T/T_F\sim 1$).
 Figure \ref{fig10} confirms the excellent agreement between the present and Becker et al. (2014) EOS's, as well as with the CEIMC (Hu et al. 2011) simulations at higher density, and the previously identified shortcomings of the SCvH EOS in this regime.
 For sake of completeness, we have carried out similar comparisons for the pressure
with the calculations of Militzer \& Ceperley (2001) for the various isochores given 
in these simulations. This is portrayed in Figure \ref{fig11}.
As seen in Fig. \ref{fig10}, we note some small wiggles in both the present and Becker et al. EOS's for the
$T=1000$ K and 2000 K isotherms (bottom curves), which stem from the imperfect interpolation procedures in their construction.

\begin{figure}
\includegraphics[width=\linewidth]{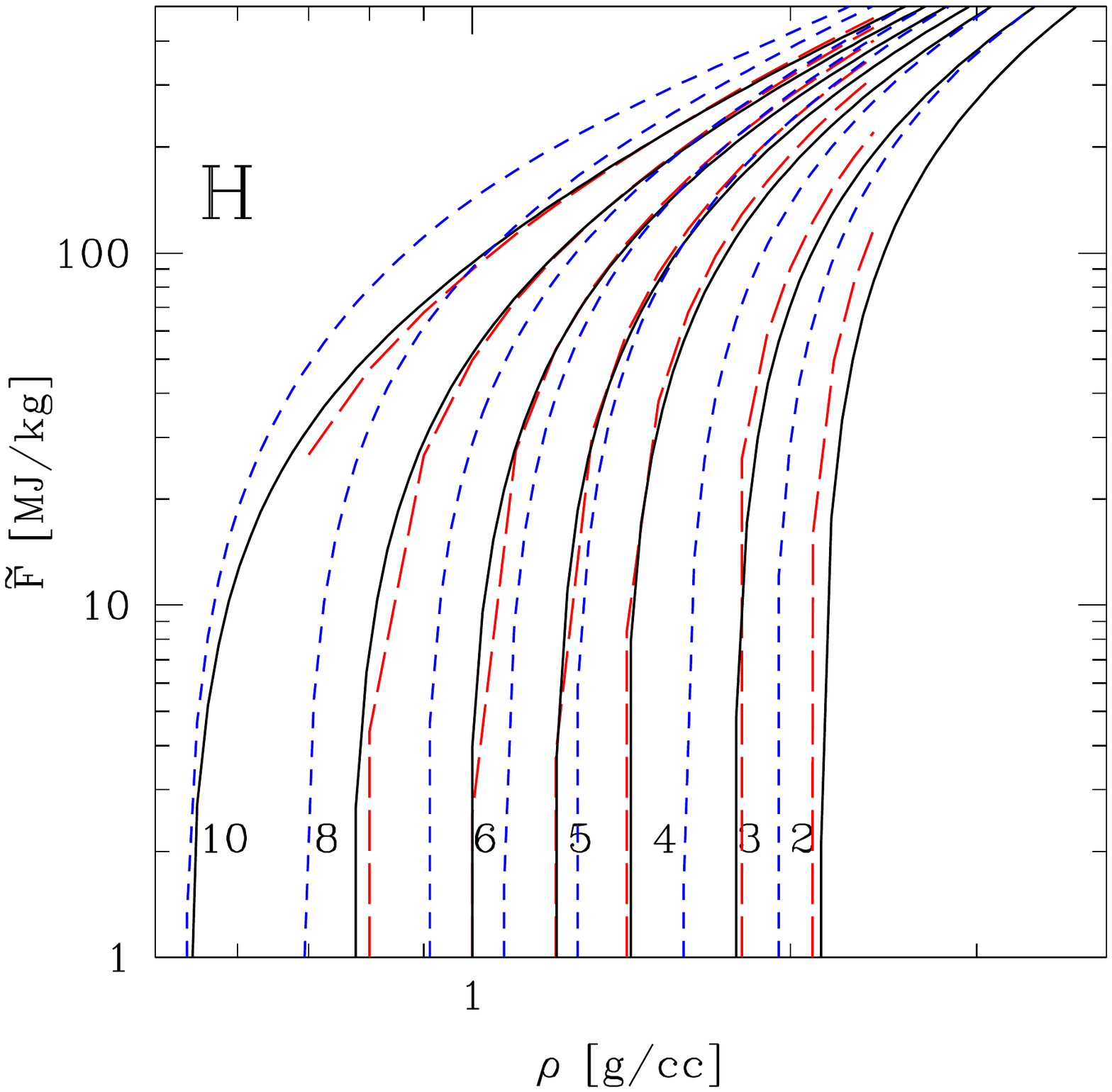}
\vspace{-2cm}
\caption{Specific free energy as a function of density for hydrogen for several isotherms (labeled in kK along the curves). 
Red long-dashed line: fit of Morales et al. (2010a); black solid line: present calculations; blue short-dashed line: SCvH.}
\label{fig12}
\end{figure}

\begin{figure}
\includegraphics[width=\linewidth]{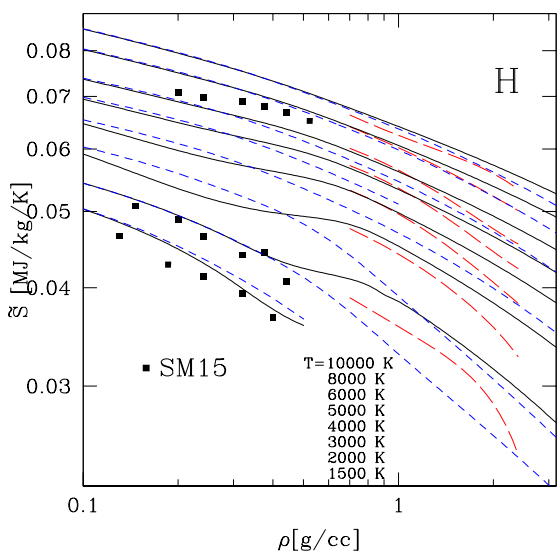}
\vspace{-2cm}
\caption{Same as Fig. \ref{fig12} for the specific entropy. Solid squares: Soubiran \& Militzer (2015, SM15).} 
\label{gig13}
\end{figure}

As mentioned in the Introduction, a major improvement of the present EOS over previous calculations is that it provides the entropy. Indeed, Caillabet et al. (2011)
derived a parameterization of the free energy $F$, yielding the entropy as $S=(F-U)/T$. 
However, as mentioned previously, the analytical fit of Caillabet et al. is valid over a rather limited temperature-density range, close to the hydrogen melting curve. In order to extend the validity of the free energy and entropy over a larger range, we have corrected the fit in various 
$T$-$\rho$ places to recover the results of Morales et al. (2010a) for H and Militzer \& Hubbard (2013) for the H/He mixture (see \S\ref{EOSHHe}).
The comparisons between the present calculations and the fitting parameterization 
derived by Morales et al. (2010a) from their simulations for
the free energy and the entropy are illustrated in Figures \ref{fig12} and
\ref{fig13}, respectively. 
Data points from QMD simulations by Soubiran \& Militzer (2015) for the entropy are also shown
in Fig. \ref{fig13}.
The agreement between the present EOS and the Morales et al. results for $F$ is excellent, in contrast to the SCvH results.
For the entropy, although not perfect, the agreement between the present calculations and the 
results fitted from the simulations is also much better than for the SCvH EOS, in particular for the coolest isotherms. 
The sudden rise of entropy above $\ga 0.3\gcc$ reflects the pressure dissociation and ionization, leading notably to an increase in
the number of particles. 
The most noticeable differences appear for the $T=6000$ K isotherm between the present (or SCvH) results and the
Soubiran \& Militzer (2015) ones in the range $\sim 0.3$-$0.6\gcc$. The discrepancy, however, remains modest,
at less than $5\%$. For the coolest isotherms, $T\le 3000$ K, the difference between the present calculations and the MC01 or SM ones rather stems from the onset of ion quantum effects (see Fig. \ref{fig1}), which are
included throughout the $\hbar^2$
Wigner-Kirkwood correction in the present calculations but are not taken into account in the simulations. 

\begin{figure}
\includegraphics[width=\linewidth]{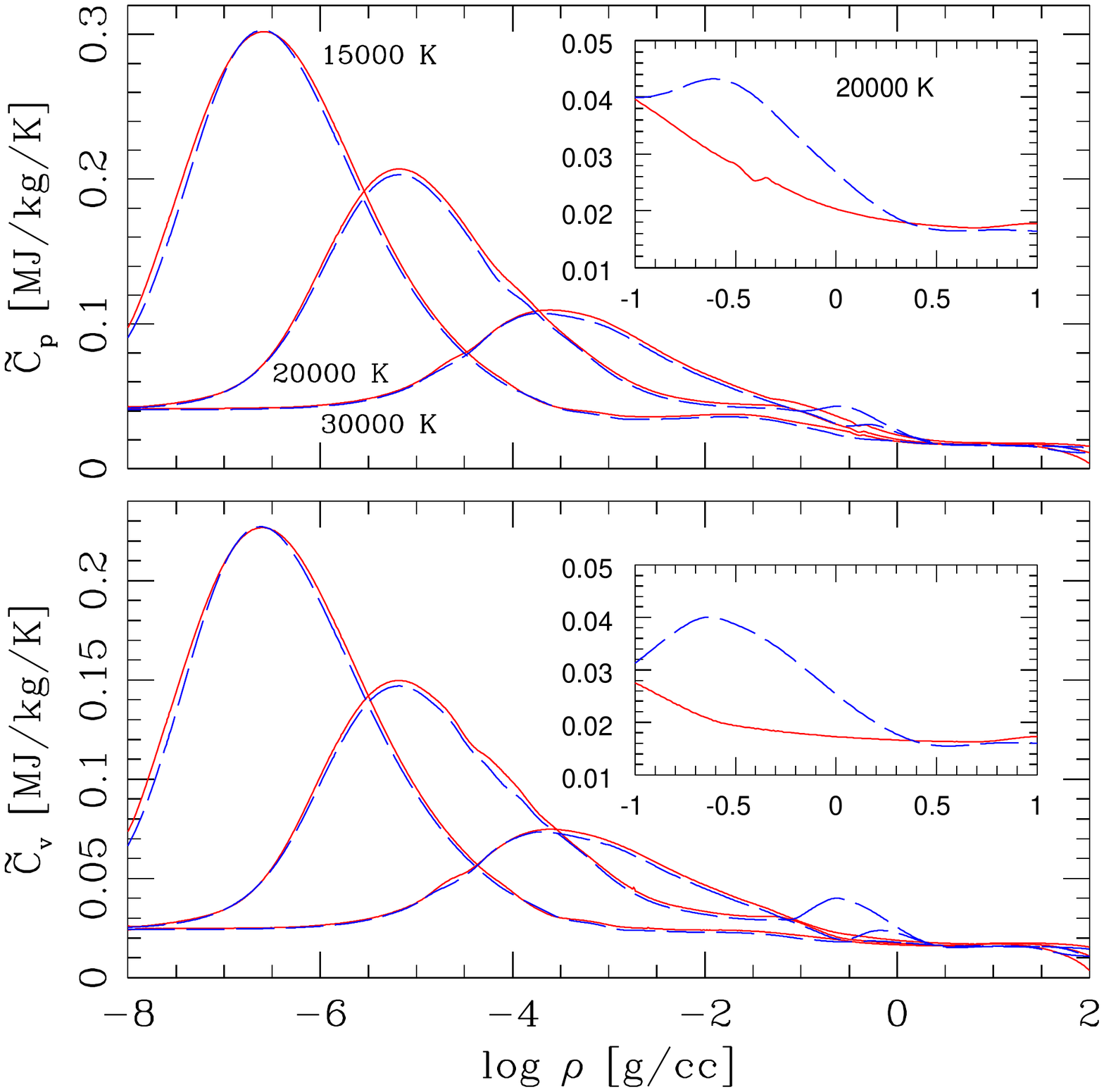}
\vspace{-2cm}
\caption{Specific heats at constant pressure and constant volume as a function of density for a few isotherms  for hydrogen.
The inset highlights the pressure dissociation/ionization domain for $T=20000$ K. Solid red: present; blue long-dashed: SCvH.} 
\label{fig14}
\end{figure}

Besides the specific internal energy $\ubar$ and entropy $\sbar$, 
the present EOS delivers all the necessary thermodynamic quantities.
These include the specific heats at constant volume and pressure, $\ucv$, $\ucp$, from the relations:

\begin{eqnarray}
\ucp &=&\sbar (\frac{\partial \log \sbar}{\partial \log T})_P \nonumber \\
\ucv &=&\sbar (\frac{\partial \log \sbar}{\partial \log T})_\rho=\ucp-\frac{P}{\rho T}\frac{\chi_T^2}{\chi_\rho}, 
\label{cpcv}
\end{eqnarray}
where $\chi_T=(\frac{\partial \log P}{\partial \log T})_\rho=-(\frac{\partial \log \rho}{\partial \log T})_P/
(\frac{\partial \log \rho}{\partial \log P})_T$ and $\chi_\rho=(\frac{\partial \log P}{\partial \log \rho})_T$. 
Figure \ref{fig14} compares these quantities for the present calculations and the SCvH ones for $T=$15000, 20000 and 30000 K. 
In the low density limit, we recover the value for molecular hydrogen with two rotational and two vibrational degrees of freedom (the rotation and vibration temperatures of H$_2$ are $\theta_{rot}=85$ K and
$\theta_{vib}=6120$ K, respectively), i.e. ${\tilde C}_V=(\frac{7}{2})\times8.25\times 10^{-3}$ MJ kg$^{-1}$ K$^{-1}$ = 0.029 MJ kg$^{-1}$ K$^{-1}$, and ${\tilde C}_P=\frac{9}{7}\times{\tilde C}_V$ = 0.038 MJ kg$^{-1}$ K$^{-1}$. As molecular dissociation or ionization
take place, the increase of the number of particles and the release of dissociation or ionization energy yield an increase of the
specific heats, with maxima corresponding to the partial dissociation or ionization zones. Eventually, at high density,
the specific heats decrease to reach the limit of a monatomic (ionized) gas, i.e. ${\tilde C}_V=\frac{3}{2}\times8.25\times 10^{-3}$ MJ kg$^{-1}$ K$^{-1}$ = 0.012 MJ kg$^{-1}$ K$^{-1}$, ${\tilde C}_P=\frac{5}{3}{\tilde C}_V$ for an ideal gas, potentially corrected by non-ideal contributions. 
As seen in the figure, these isotherms bracket the domain of
 hydrogen {\it pressure} dissociation and ionization which occurs between about $\sim 0.1$ and 3.0 $\gcc$. The inset clearly highlights the lack of H$_2$
 ionization at high density in the SCvH model.

\begin{figure}
\includegraphics[width=\linewidth]{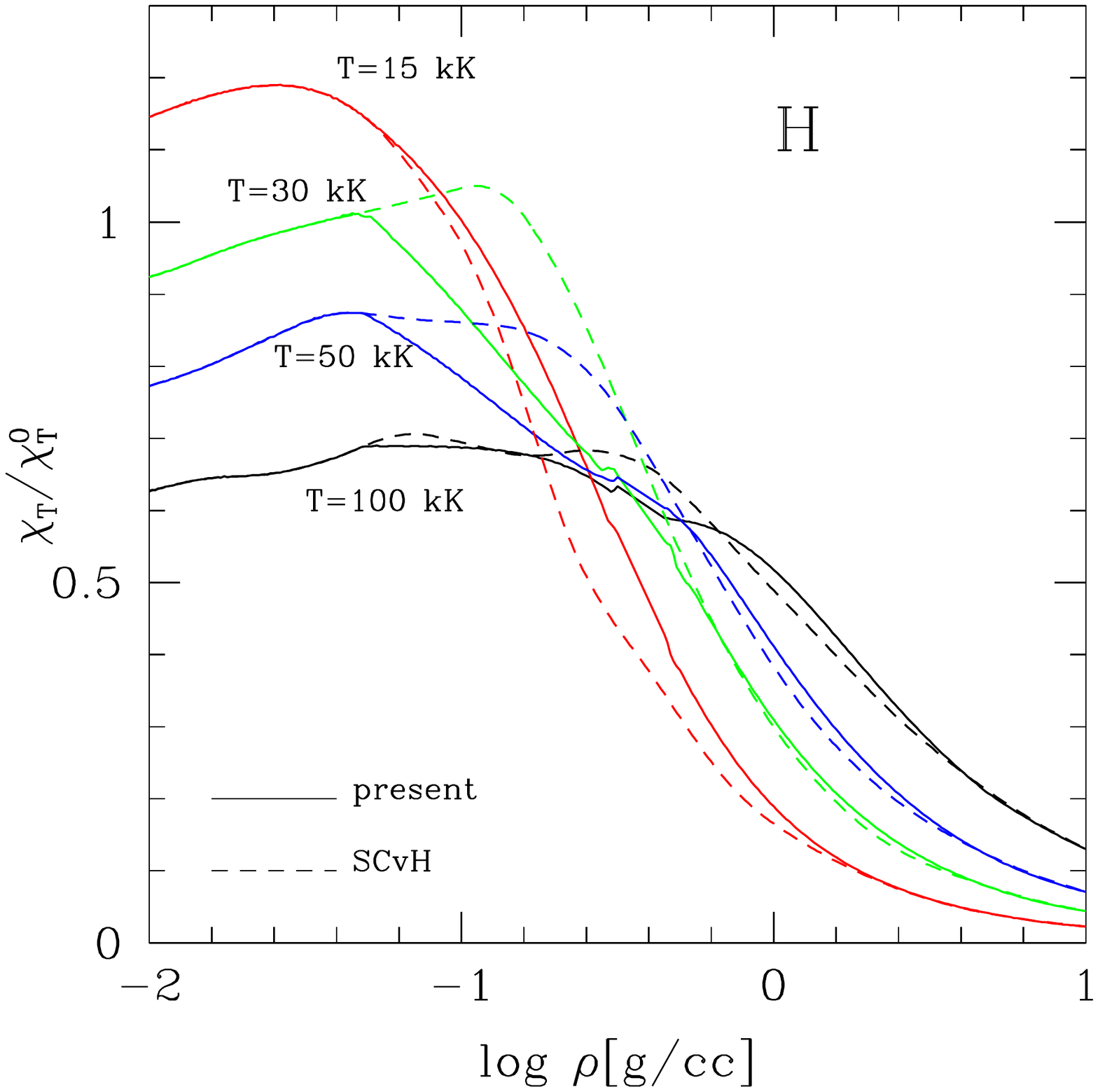}
\vspace{-2cm}
\caption{Isothermal compression factor as a function of density  for hydrogen for a few isotherms (as labeled in kK) for the
present (solid) and SCvH (dash) calculations.
The compressibilities are normalized to the one of a perfect monatomic H gas.} 
\label{fig15}
\end{figure}

Figure \ref{fig15} displays the isothermal compressibility factor 

\begin{eqnarray}
{\kappa_T\over \kappa_T^0}={\rho \kb T\over \mu P}\chi_\rho^{-1},
\label{chi}
\end{eqnarray}
 where $\kappa_T^0=m_H/(\rho \kb T)$ is the isothermal compressibility of a perfect
monatomic hydrogen gas and $\mu=A\mh$ is the atomic weight, with $\mh=1.660\times 10^{-27}$ kg the atomic unit mass, for the present and SCvH calculations over the temperature and density domain characteristic of
hydrogen pressure and thermal dissociation and ionization. 
The figure highlights the too large compressibility of the SCvH EOS in this
domain, as noticed in \S2.2 over the Hugoniot compression curves, due to the lack of H$_2$ dissociation/ionization.


\section{The Helium equation of state}
\label{EOSHe}
\subsection{Construction of the EOS model}
\label{Hemodel}

The procedure for the helium EOS is exactly the same as for the hydrogen one, with the combination of
 different calculations. 
For $T\ge 1.0\times 10^6$ K, the plasma becomes fully ionized and we use the Chabrier \& Potekhin (1998) EOS model. 
For  $T< 1.0\times 10^6$ K, the EOS is divided again into 3 density regimes.
In the low density (atomic) one, we use the SCvH EOS for pure He.
In the intermediate $T$-$\rho$ regime, the ab-initio calculations are based on original QMD simulations and will be described in details in a dedicated paper (Soubiran et al. 2019, in prep.). 
In the high density, fully ionized domain, we use the Chabrier-Potekhin (1998) EOS.

\begin{itemize}
\item $\rho \le 0.1\,\gcc$: SCvH EOS
\item $1.0< \rho\le 100.0\,\gcc$: EOS of Soubiran et al. (2019), based on QMD calculations
\item $\rho> 100.0\,\gcc$: CP98 EOS
\end{itemize}
As for hydrogen, bicubic spline procedures are used to interpolate the thermodynamic quantities in the intermediate regime.
The fact that we merge the QMD and CP98 calculations at $100\gcc$ is justified by the fact that QMD calculations (Soubiran et al. 2012), reanalyzing reflectivity measurements of dense fluid helium (Celliers et al. 2010) by
including the effects of temperature on the helium gap, suggest that this latter closes at a density of about 10$\gcc$, in good
agreement with previous semi-analytical models (Winisdoerffer \& Chabrier 1995), implying that
helium should be fully ionized above this density. The zero of energy for the helium EOS is the same as in SCvH, namely the zero of the isolated He atom.

\begin{figure}
\includegraphics[width=\linewidth]{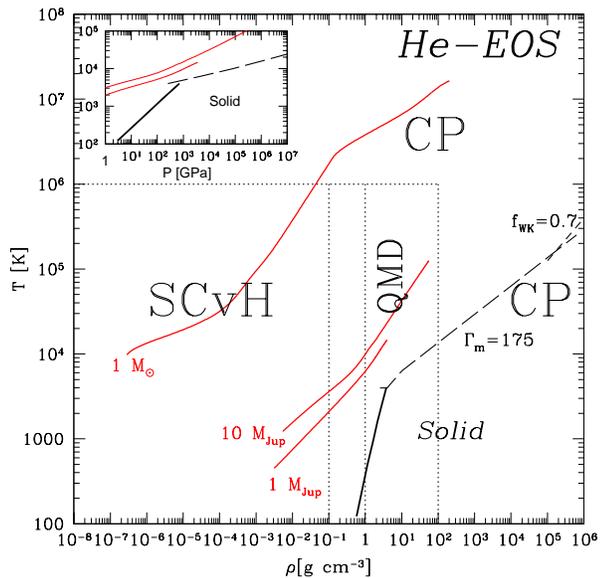}
\vspace{-2cm}
\caption{Temperature-density domain of the present EOS for helium. The dotted
line gives the $T$-$\rho$ domains corresponding to the different models or calculations combined to produce the final EOS (see text). 
The melting lines for He (eq. (\ref{Hemelt})) and
He$^{2+}$ (eq. (\ref{OCP})) are delimited by the solid and long-dashed lines, respectively, in the lower right
corner (note that the line for He$^{2+}$ is extrapolated beyond the validity of the OCP model for illustrative purposes only). The short-dash line $f_{WK}=0.7$ corresponds to the limit of validity of the present calculations, due to ion
quantum effects. The inset focusses on the liquid to solid and ion classical
to quantum locations of the phase diagram in $T$-$P$. The EOS must not be used beyond these limits. 
Interior profiles for the Sun (1 $\msol$) and 1 and 10 $\mjup$ planets
at 5 Gyr (from Baraffe et al. 2003, 2015) are displayed on the figure to illustrate the domain of astrophysical applications. }
\label{fig16}
\end{figure}

As for hydrogen, although, for practical purposes, the tables are calculated over square $T$-$\rho$ and $T$-$P$ domains, part of these
latter are meaningless, as they correspond to regions where either helium becomes solid (Loubeyre et al. 1993) or quantum diffraction effects for ions become
dominant. 
The melting line for helium, determined by diamond anvil cell experiments, is well described by a simple Simon law (Datchi et al. 2000), even when extrapolated to Mbar pressures, where this expression is in good
agreement with ab initio calculations (Lorenzen et al. 2009):

\begin{eqnarray}
T_m=61.0\,P^{0.639}\,\,{\rm K},
\label{Hemelt}
\end{eqnarray}
where the pressure $P$ is in kbar ($=0.1$ GPa). This is indicated by the thick solid line in Fig. \ref{fig16}, while the OCP melting line (eqn.(\ref{OCP})) is shown by the long-dashed line. Interestingly enough,
as for hydrogen, we see that this curve, when extended to low temperatures and densities, nicely merges with the experimental one.
As for the onset of quantum diffraction effects, the characteristic parameter is the same as for hydrogen, $f_{WK}\ga 0.7$, and the condition given by eqn.(\ref{fwk}) is indicated by the short-dashed line.

\subsection{Comparison with ab initio calculations}

\begin{figure}
\includegraphics[width=\linewidth]{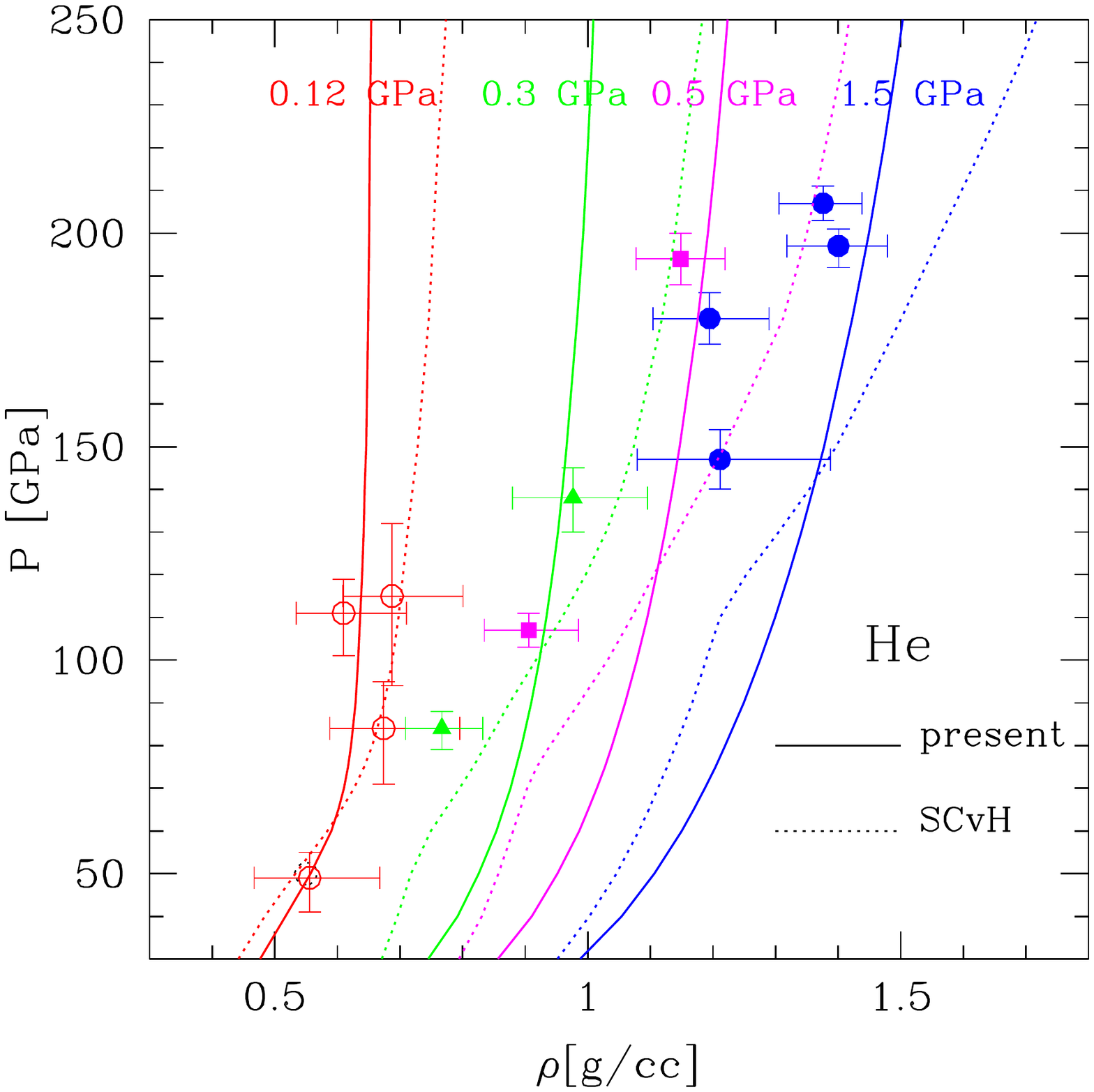}
\vspace{-2cm}
\caption{Helium shock pressure vs density along the Hugoniot for various precompressed
initial conditions, namely 0.12, 0.3, 0.5 and 1.1 GPa, as labeled in the figure. Data: Brygoo et al. (2015); solid line: present calculations; dotted line: SCvH EOS.}
\label{fig17}
\end{figure}

\begin{figure}
\includegraphics[width=\linewidth]{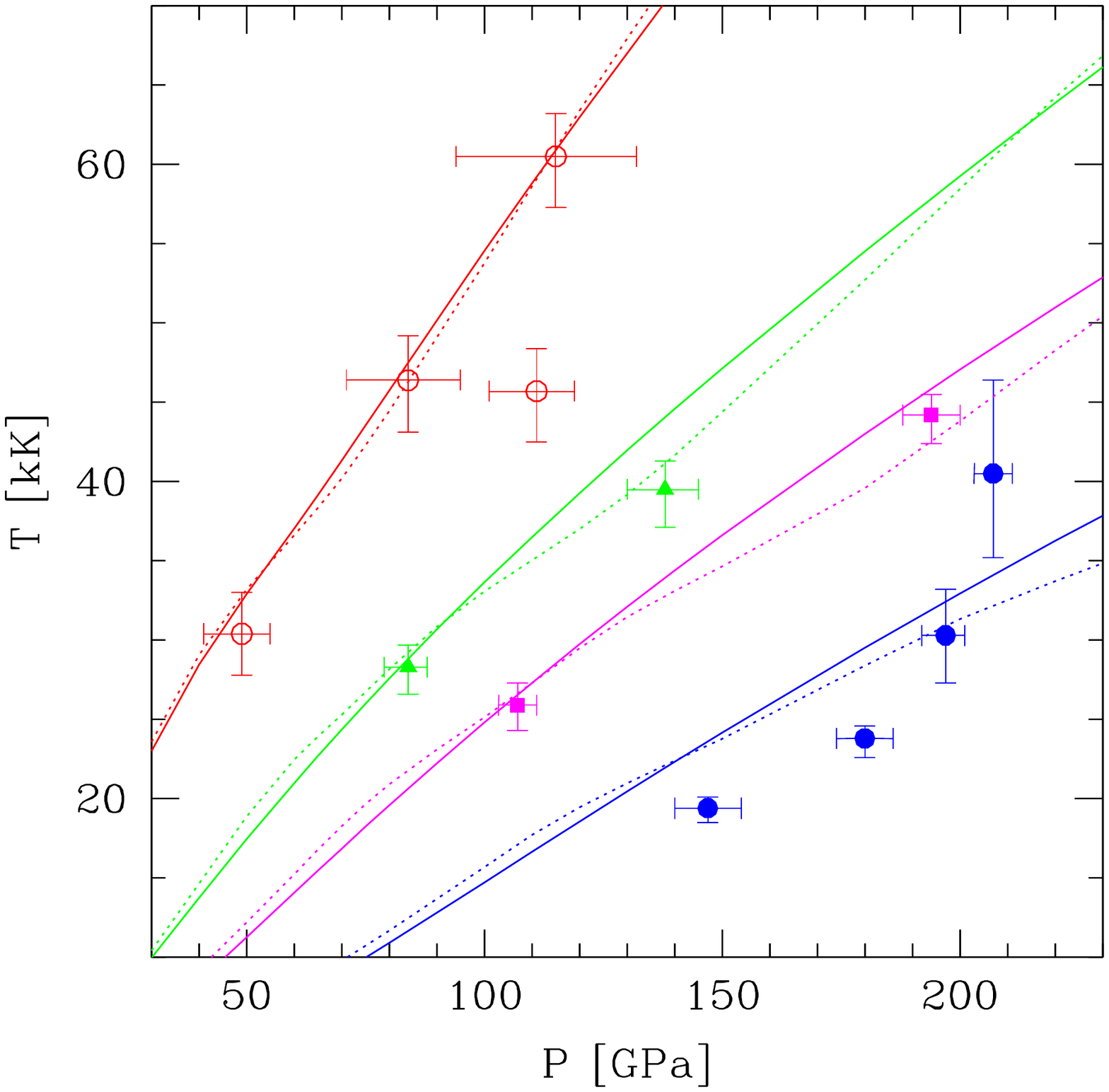}
\vspace{-2cm}
\caption{Helium shock temperature as a function of shock pressure along the Hugoniot for the same precompressed
initial conditions as in Fig. \ref{fig17}, namely 0.1, 0.3, 0.7 and 1.5 GPa from left to right. Same labeling as in Fig. \ref{fig17}.}
\label{fig18}
\end{figure}

Extensive comparisons between this pure He EOS and PIMC or existing QMD simulations will be
presented in details in a forthcoming paper (Soubiran et al., 2019, in prep.). Meanwhile,
Figures \ref{fig17} and \ref{fig18} compare the Hugoniot 
compression curves obtained with our He EOS with the recent data of Brygoo et al. (2015), 
for different precompressed initial conditions.
The agreement between the present calculations and the data is very good, except for the two
data points at lowest pressure of the $P_0=0.3$ and 0.5 GPa precompressed experiments, which
seem to be surpringly stiff.
As for hydrogen, we note that the SCvH model predicts a too abrupt ionization compared with
both the experiments and the present calculations, which rather suggest a smoothly ongoing
process. This may again point to a limitation of the so-called chemical semi-analytical model, based on
the concept of pair potentials for the various species (at present He, He$^+$ and He$^{++}$),
to describe N-body interactions, as the same abrupt ionization is found in the more sophisticated model of Winisdoerffer \& Chabrier (2005) for helium.

\section{The Hydrogen/Helium mixture equation of state}
\label{EOSHHe}

\subsection{Calculation of the H/He EOS}

The calculations of the EOS for the H/He mixture are carried out 
within the so-called "additive volume law" (AVL), as in SCvH. This latter is based on the additivity of the
extensive variables (volume, energy, entropy,...) at constant intensive variables (P,T). Although this method is formally exact for non-interacting, ideal mixtures, 
and excellent in the limit of fully ionized systems (Chabrier \& Ashcroft 1990), it
is no longer the case for interacting systems, i.e. between hydrogen and helium species in the present context, or in the domain of partial ionization. Nevertheless, we expect the correction to
remain modest, of the order of a few per cent. Clearly, this is a limitation of the present EOS. Calculations for the interacting H/He mixture have been
carried out recently with QMD simulations but have focused on a limited density-temperature domain characteristic of Jupiter internal adiabat (Morales et al. 2010b, Lorenzen et al. 2009, 2011, Militzer 2013, Mazzola et al. 2018, Sch\"ottler \& Redmer 2018) and only one of these calculations has calculated the entropy of the interacting mixture (Militzer \& Hubbard 2013). Note also that
 all these simulations have been carried out with a rather small number of particles,
of the order of 10 for the helium atoms. Indeed, for large numbers of particles, demixing can occur in the simulation box, preventing the calculation of the thermodynamic properties of the mixture (Lorenzen et al. 2009, Soubiran et al. 2013).
 Therefore, even though various schemes exist to correct for the finite size errors, 
it seems fair to say that the quantities derived from existing simulations still retain some degree of uncertainty.

As mentioned above, within the AVL, an extensive variable $W$ at given $(T,P)$ for the mixture reads (see, e.g. SCvH):

\begin{eqnarray}
W(T,P)=\sum_i X_i W_i(T,P),
\label{W}
\end{eqnarray}
where $X_i=M_i/(\sum_i M_i)$ denotes the mass fraction of component $i$ (H or He in the present context). The density for the H/He mixture (which is an inverse specific volume) reads

\begin{eqnarray}
{1\over \rho(T,P)}={1-Y\over \rho_H(T,P)} + {Y\over \rho_{He}(T,P)},
\label{dens}
\end{eqnarray}
where $Y=M_{He}/(M_H+M_{He})$ denotes the helium mass fraction. For the specific entropy of the mixture, the ideal mixing entropy must be added to eqn.(\ref{W}) in order to correctly recover the ideal gas limit, yielding

\begin{eqnarray}
{\tilde S}(T,P)=\sum_i X_i {\tilde S_i}(T,P) + {\tilde S}^{id}_{mix}(T,P).
\label{Smix}
\end{eqnarray}
For a mixture of $N=\sum_i N_i$ components $i$ of number fraction $x_i=N_i/N$ and atomic mass $A_i m_H$, with $m_H$ the atomic mass unit, the ideal mixing entropy reads
\begin{eqnarray}
{{\tilde S}^{id}_{mix}\over k_B} &=& \frac{1}{N\, \langle A\rangle}(N\ln N-\sum_i N_i\ln N_i)\nonumber \\
&=& -{ x_H\ln \,x_H + x_{He} \ln \,x_{He} \over {\langle A\rangle} \mh},
\label{Sid}
\end{eqnarray}
where $\langle A\rangle=\sum_i x_i A_i$ and $k_B$ denotes the Boltzmann constant. It should be noted that in the above equation, we have neglected the
contribution from the free electron entropy. Indeed, in contrast to semi-analytical so-called "chemical models" such as SCvH, based on well defined chemical entities such as molecules, atoms and electrons, such an identification does not exist in QMD simulations, preventing the precise caracterization of a free electron density.
Our approximation, however, is justified both in the regime of neutral hydrogen and helium, where there are no free electrons, and in the regime of full ionization, where
the electrons become degenerate and thus have a negligible entropy. 
The approximation, however, fails in between these two regimes, i.e. in the regime of partial ionization.

\begin{figure}
\includegraphics[width=\linewidth]{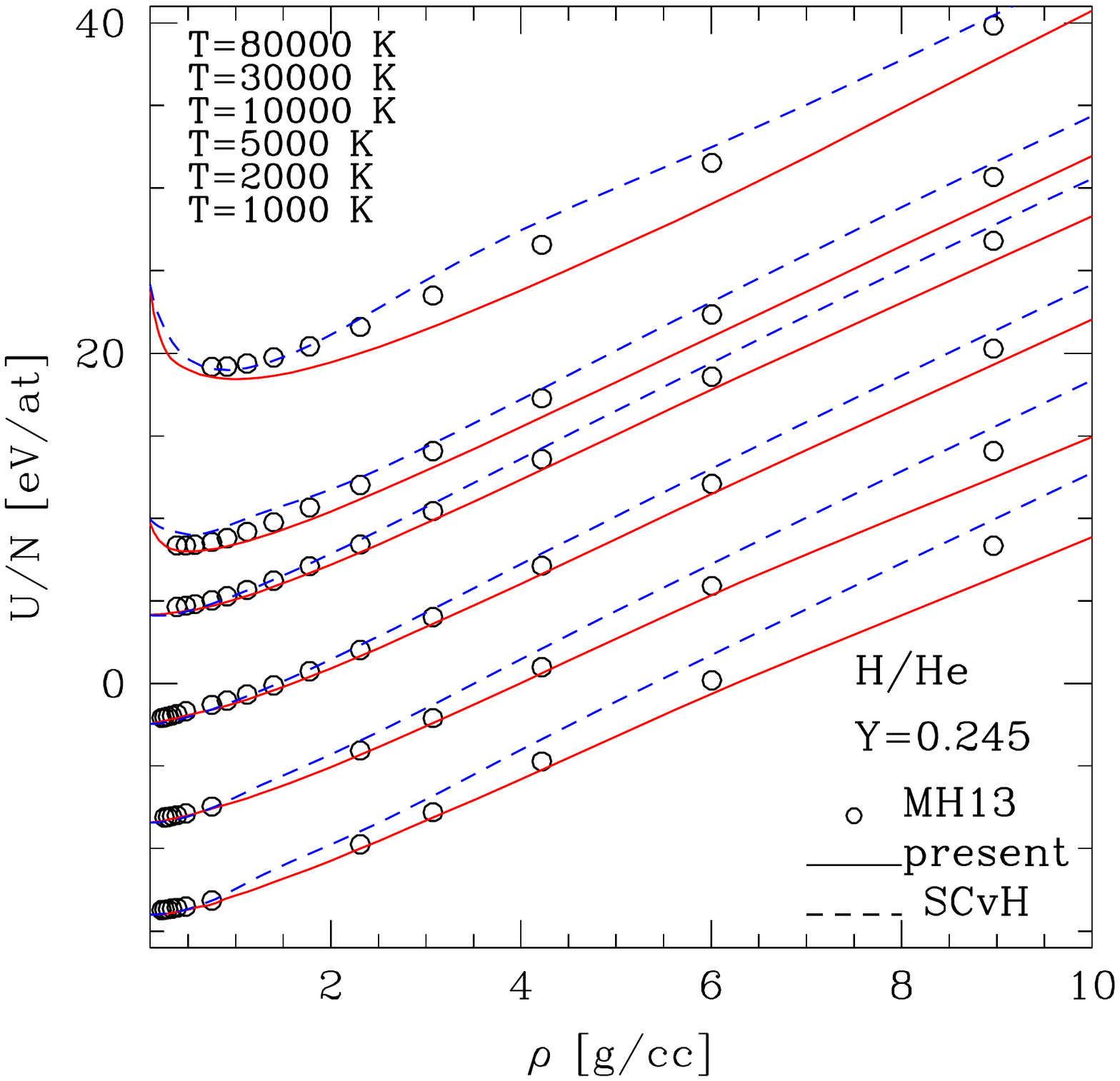}
\vspace{-2cm}
\caption{Internal energy per atom vs density for several isotherm calculations by Militzer \& Hubbard (2013, MH13) (as labeled in the figure), compared with the present and SCvH results, respectively.
For all curves the zero of energy is the same as in MH13.
For sake of clarity, however, curves have been arbitrarily moved upward or downward by constant shifts.} 
\label{fig19}
\end{figure}

\begin{figure}
\includegraphics[width=\linewidth]{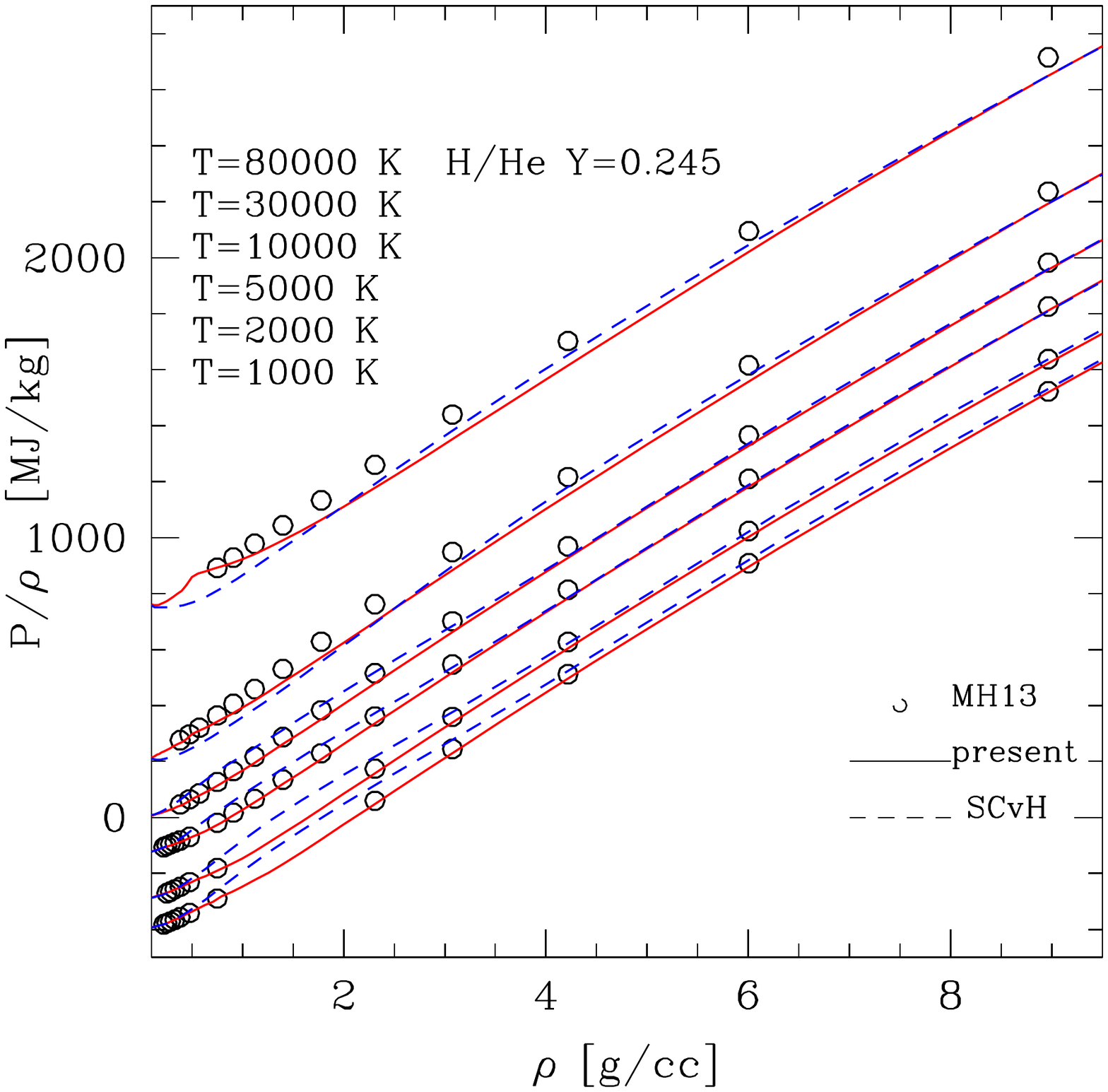}
\vspace{-2cm}
\caption{Same as Fig. \ref{fig19} for the non-ideal pressure $P/\rho$.} 
\label{fig20}
\end{figure}

As mentioned above, Militzer (2013) and Militzer \& Hubbard (2013) have carried out QMD simulations for a H/He mixture over a significant temperature-density domain and have calculated the free energy $F$ by thermodynamic integration, which yields also the entropy $S=(U-F)/T$. These authors provide a fitting formula for $F$ over the range of their
simulations, namely $\sim 0.2$-9.0$\gcc$ and 1000-80000 K, covering the domain of H and He pressure and temperature dissociation and ionization.
The simulations were carried out for a helium number fraction $x_{He}=18/238=0.076$, corresponding to a helium mass fraction $Y=0.246$. 
Figures \ref{fig19} and \ref{fig20} display the comparison of the internal energy $E$ 
and the excess pressure $P/\rho$ as a function of density for several isotherms calculated by MH13 with the present and SCvH results. As noted in MH13, the SCvH EOS generally
slightly overestimates the internal energy compared with the simulations over the probed density range. This is improved with the new EOS, although this latter predicts a {\it lower} internal energy than MH13 for the highest isotherms, most likely due to the temperature and density
interpolation procedures between the QMD and CP98 calculations in this domain. 
We also have to recall that non-ideal H/He mixing effects are not included in the present calculations, based on the additive volume law. Indeed, it has been shown that in the ionized regime, this latter yields a lower energy than for the non-ideal mixture (Chabrier \& Ashcroft 1990).
The sharp increase in the energy in the density regime $\sim 0.5$-1.0 $\gcc$ in the displayed temperature regime stems from the onset of ionization, yielding a strong increase of the free electron 
energy contribution due to the Pauli principle (see e.g. Saumon \& Chabrier 1992, Militzer \& Hubbard 2013).

\begin{figure}
\includegraphics[width=\linewidth]{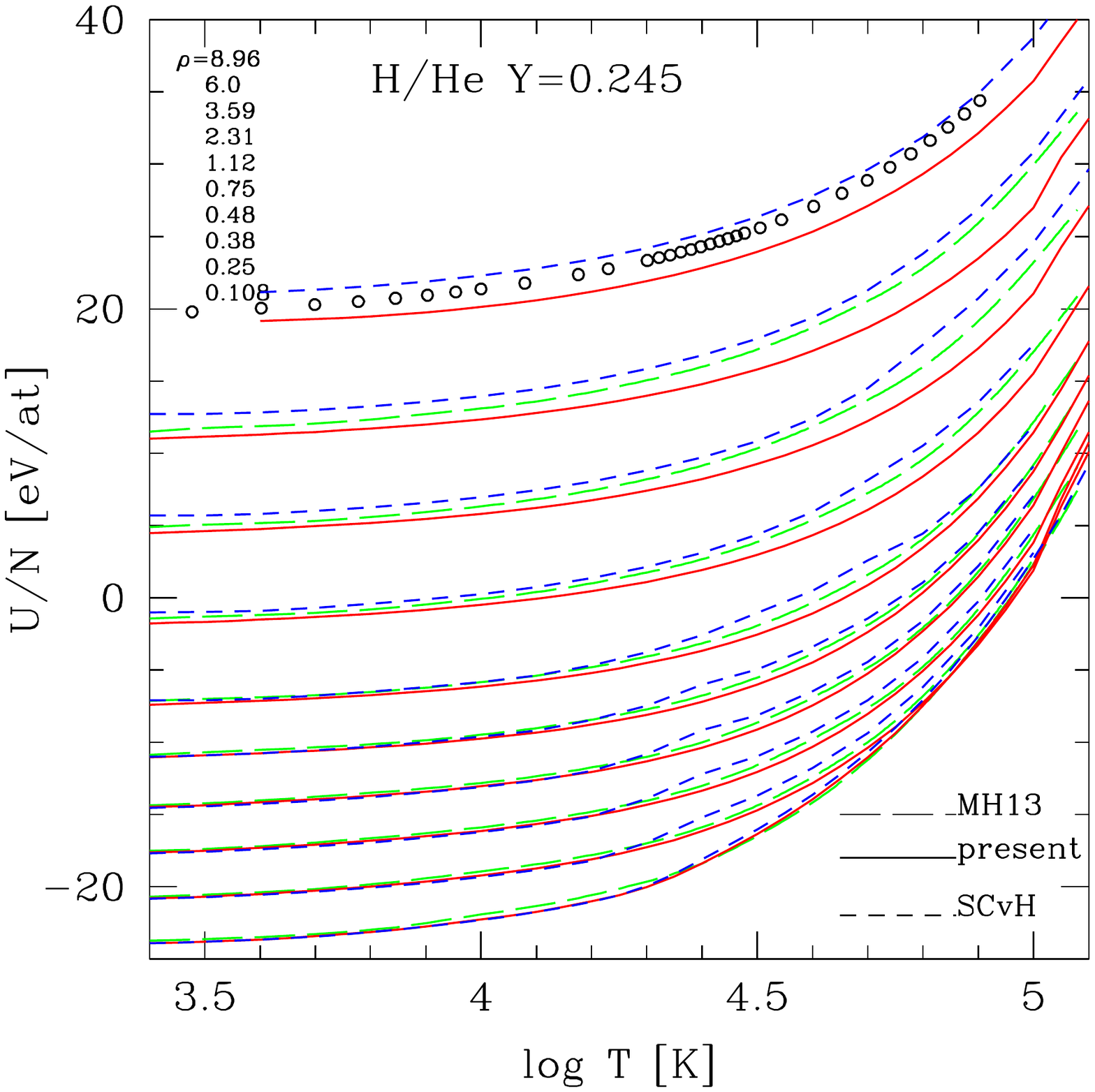}
\vspace{-2cm}
\caption{Internal energy per atom vs temperature for several isochore calculations by Militzer \& Hubbard (2013, MH13) (as labeled in the figure), compared with the present and SCvH results.
For all densities, the MH13 values are the ones given by their fit except for $\rho=8.96\gcc$, which is out the range of validity
of the fit and for which the empty circles are their simulation data points.
Solid lines: present calculations; green long-dashed lines and empty circles: MH13; blue short-dashed lines: SCvH.
For all curves the zero of energy is the same as in MH13.
For the sake of clarity, however, curves have been moved arbitrarily upward or downward by constant shifts.} 
\label{fig21}
\end{figure}

\begin{figure}
\includegraphics[width=\linewidth]{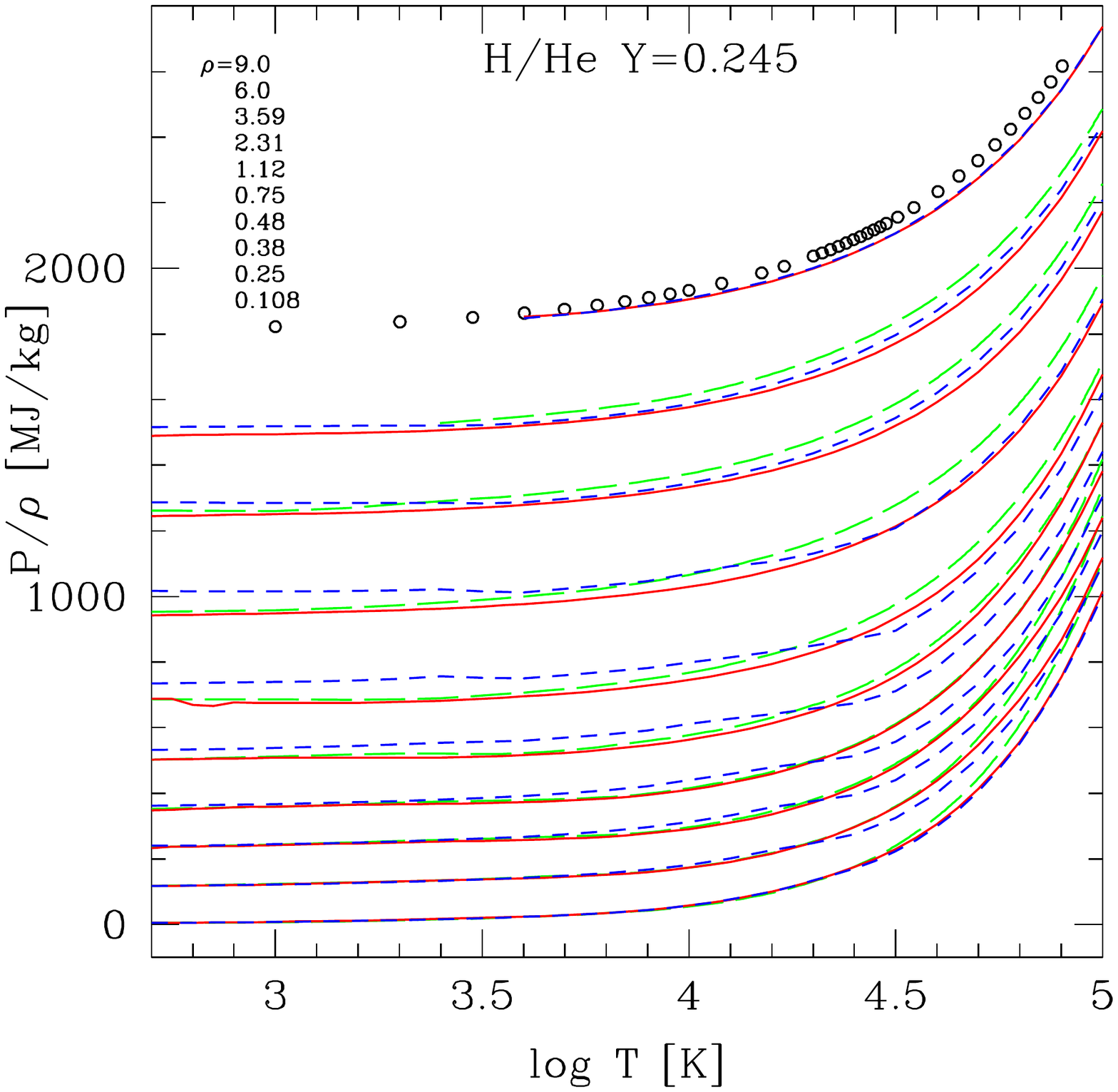}
\vspace{-2cm}
\caption{Same as Fig. \ref{fig21} for the non-ideal pressure $P/\rho$. 
 For the sake of clarity, curves have been shifted upward arbitrarily by constant shifts.}
\label{fig22}
\end{figure}

Figures \ref{fig21} and \ref{fig22} portray similar comparisons as a function of temperature for several isochores calculated by MH13. In Figure \ref{fig21}, we notice the abrupt increase of the SCvH energy w.r.t. both MH13 and the present calculations around $T\sim 30000$ K and $\rho \gtrsim 0.2\gcc$, i.e.
in the regime of pressure dissociation/ionization, while the reverse is true for the pressure,
with a cross over of the SCvH isochores with the present and MH13 ones in the regime $\sim$0.2-3.0 $\gcc$ around $T\sim 30000$ K ($\log T=4.5$) (see Fig. \ref{fig22}). 
This again illustrates the approximate treatment of this process in the semi-analytical SC model, as mentioned in the previous sections and as
already noted by Militzer \& Ceperley (2001) for pure hydrogen and
Militzer \& Hubbard (2013) for the mixture, highlighting the already mentioned lack of dissociation and too abrupt ionization in SCvH with increasing density. The underestimated degree of molecular dissociation and/or ionization in the SC model is also reflected 
by the increasing offset between SCvH and both MH13 and the present calculations for
$T\lesssim 10^4$ K for both $U$ and $P/\rho$ in the density regime $\sim 0.75$ - 3.6$\gcc$.  
As pointed out by MH13, this discrepancy on the pressure as a function of temperature can have a significant impact on giant planet internal structures.
In contrast, the agreement between the present EOS and MH13 in this crucial domain can be considered as satisfactory.
At higher densities, when the system becomes dominantly ionized, all calculations agree.  
Generally speaking, the present EOS agrees well with the MH13 simulations,
except possibly in the domain $4.5\la {\rm log}\,T\la 5.0$ for for $\rho \simeq 2.0$-6.0$\gcc$, as seen in the figures, with a maximum discrepancy of $\sim 8\%$ .
Since this is within the domain of interpolation between the QMD-based simulations and the CP98 model in the present
EOS calculations (see \S\ref{Hmodel}),
the discrepancy is likely to be blamed upon this procedure.

As already mentioned, besides the pressure and the internal energy, the knowledge of the entropy is 
necessary to determine the thermal profile and the cooling rate
of objects below about 0.6 $\msol$. This domain encompasses low-mass stars, brown dwarfs and gaseous planets. Indeed, these objects are too cool for heat to be
transported efficiently by radiation and not dense enough for electron conduction to be significant. Heat is thus transported by convection, yielding a nearly adiabatic internal profile.
Deriving the entropy over a large enough temperature-density range to cover the evolution of these bodies is thus of prime importance for astrophysical applications as
 well as for isentropic high-pressure experiments aimed at characterising hydrogen and helium pressure ionization. So far, no such EOS has been derived. 
 
 Militzer (2013) and Militzer \& Hubbard (2013) calculated the Helmholtz free energy from their QMD simulation data by performing a so-called thermodynamic integration technique (TDI).
 The advantage of this technique, where integration is performed over trajectories that are derived from a hybrid potential energy function between the one of a classical system and the 
one obtained with a Kohn-Sham functional, is that it does not require a prohibitively large number of $(T,\rho)$ simulation points.
The other advantage of the TDI method is that it allows directly the determination of the ionic contributions to the entropy. Whereas in most cases this contribution
represents essentially a measure of the total entropy of the system, this is no longer true when electronic excitations become important, i.e. once ionization takes place. In that
case the electron contribution to the entropy must be taken into account in the TDI integration
(see Militzer (2013) and references therein).
It must also be kept in mind that the procedure becomes
less straightforward in the molecular regime, where a rigorous classical reference system is more difficult to define, because of exchange reactions, leading to dissociation and
recombination. Last but not least, in some (low)temperature-(high)density domain, corrections due to quantum effects in the motion of nuclei must also be taken into account in the DFT-MD results. Finally, finite-size effects due to the limited number of particles must be treated with extreme care to ensure they do not affect substantially the results.
Computational calculation of the entropy of a system is thus a highly delicate task and is not free from
uncertainties.

\begin{figure}
\includegraphics[width=\linewidth]{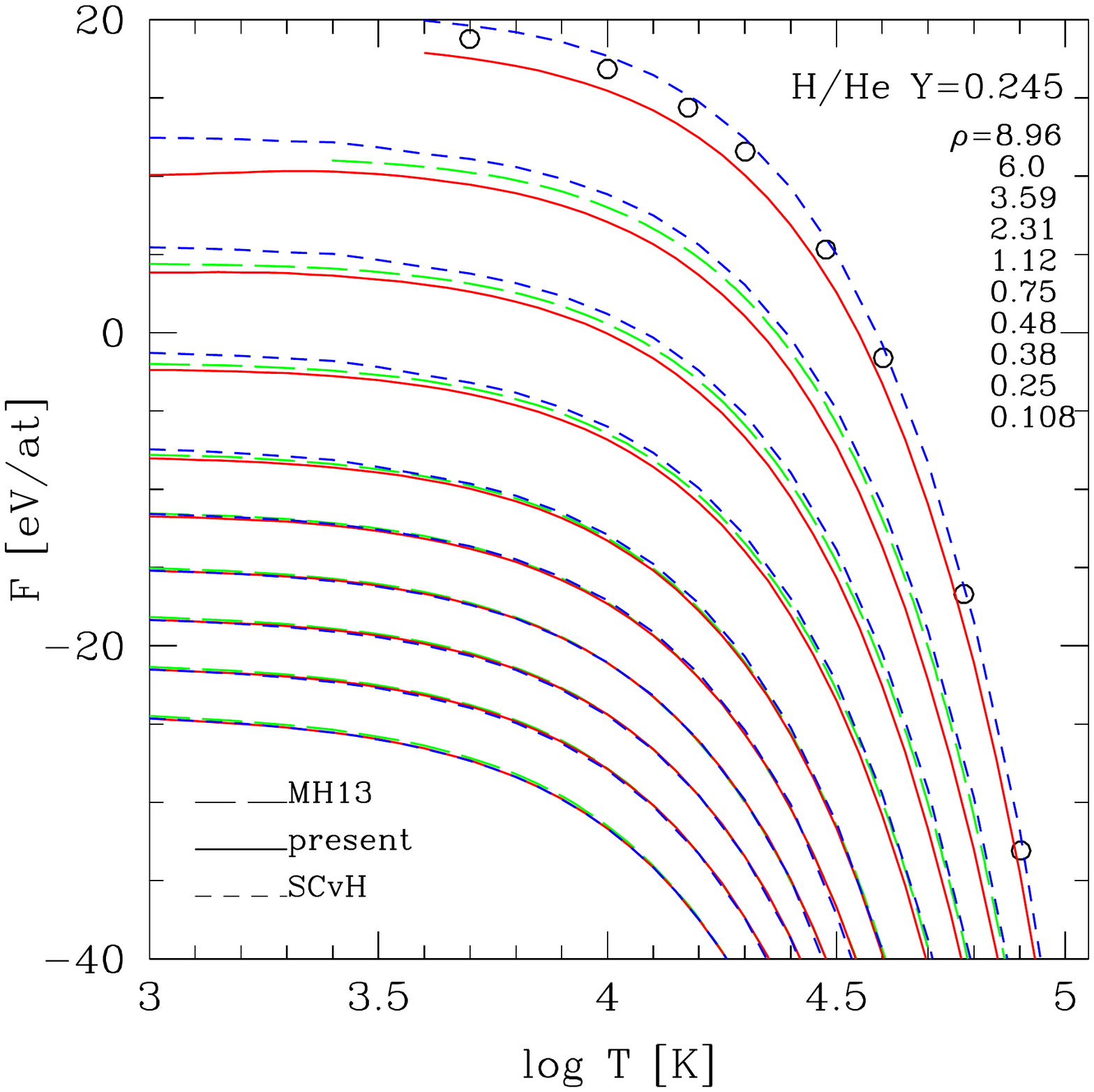}
\vspace{-2cm}
\caption{Free energy per atom vs temperature for several isochore calculations by Militzer \& Hubbard (2013) (as labeled in the figure), compared with the present and SCvH results, respectively. 
Empty circles: MH13 computation data; green long-dashed lines: MH13 fit; solid lines: present calculations; blue short-dashed lines: SCvH.
For all curves the zero of energy is the same as in MH13.
}
\label{fig23}
\end{figure}

\begin{figure}
\includegraphics[width=\linewidth]{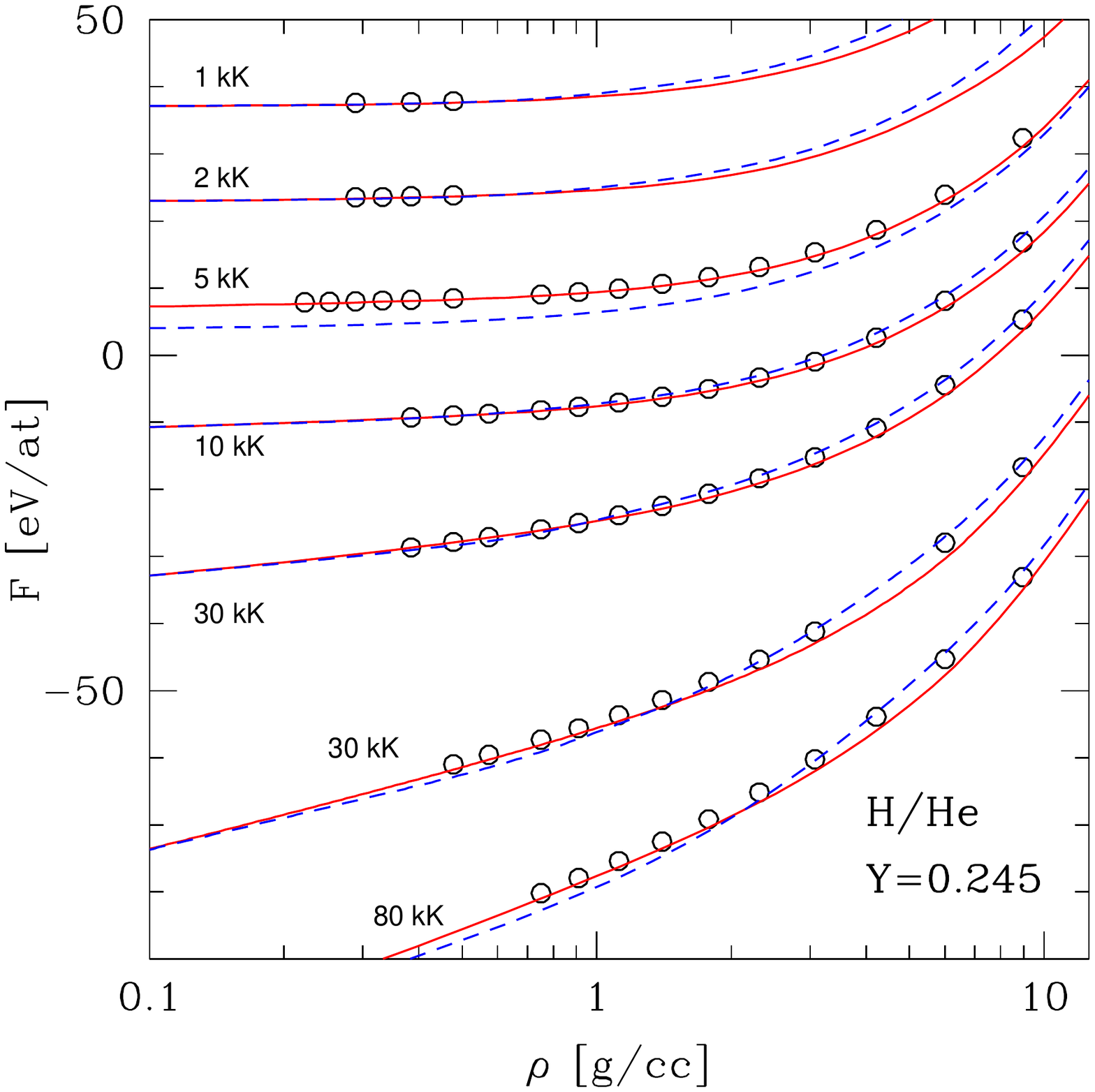}
\vspace{-2cm}
\caption{Free energy per atom vs density for several isotherm calculations by Militzer \& Hubbard (2013, MH13) (as labeled in the figure), compared with the present and SCvH results. 
For sake of clarity, however, curves have been shifted upward for the 1000 K, 2000 K and 5000 K isotherms.
Same labeling as Fig. \ref{fig23}.} 
\label{fig24}
\end{figure}

Figures \ref{fig23} and \ref{fig24} portray a comparison of the free energy $F$ 
per atom as a function of temperature and density, respectively, between the present calculations, the SCvH EOS and the MH13 simulations in the density-temperature range probed by these latter, using either their numerical data points
or their polynomial fit within its domain of validity. As already noticed by MH13, the agreement for this quantity is much better
 than for the pressure and the internal energy, which are respectively the density and temperature {\it derivatives} of $F$. We note, however, the better agreement of the present calculations with the
simulations compared with SCvH in the $T$-$\rho$ domain
 where ionization sets in.

\begin{figure}
\includegraphics[width=\linewidth]{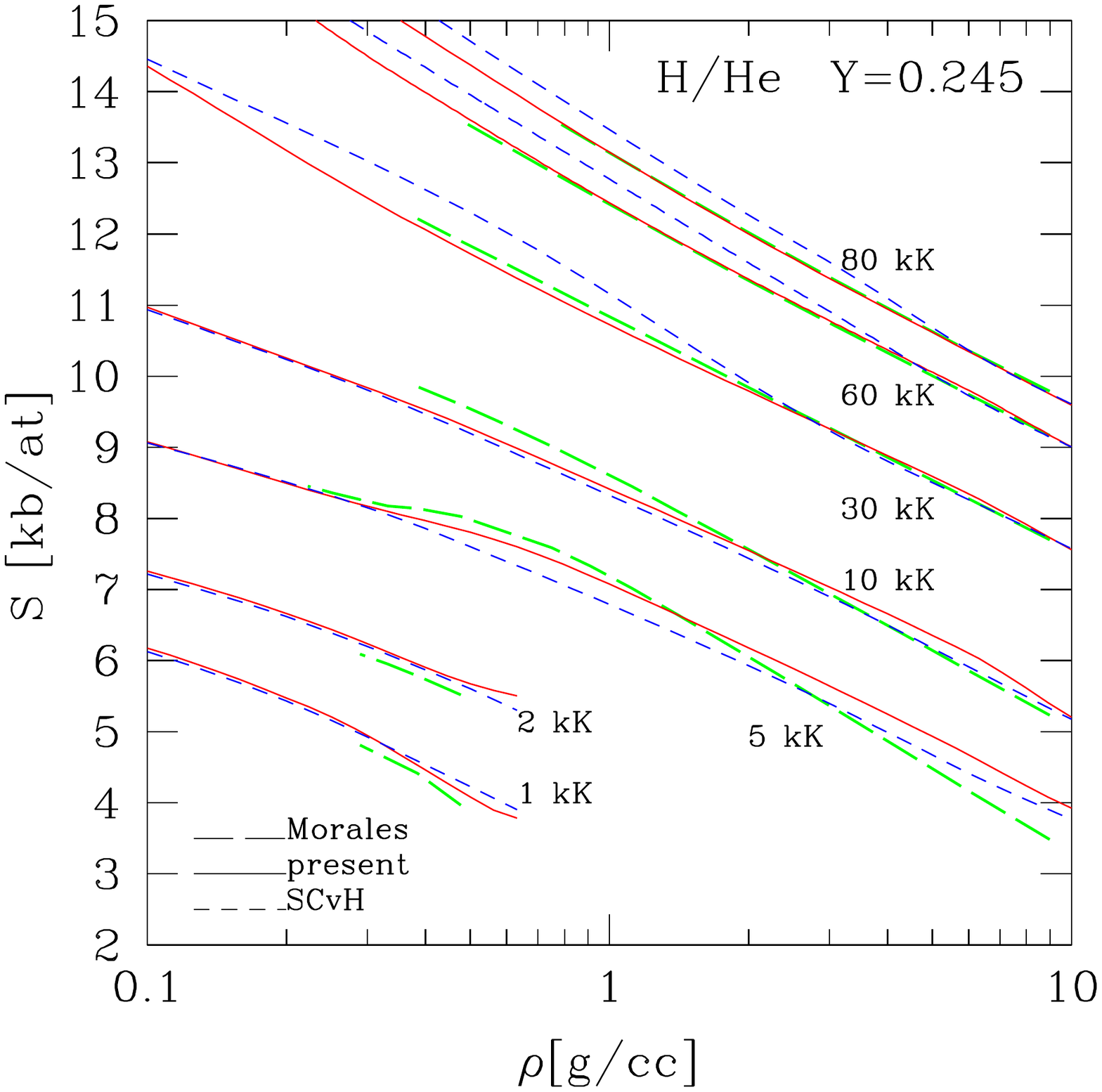}
\vspace{-2cm}
\caption{Entropy vs density for several isotherm calculations by Militzer \& Hubbard (2013) (as labeled in the figure), compared with the present and SCvH results, respectively. 
Same labeling as in the previous figures.} 
\label{fig25}
\end{figure}

\begin{figure}
\includegraphics[width=\linewidth]{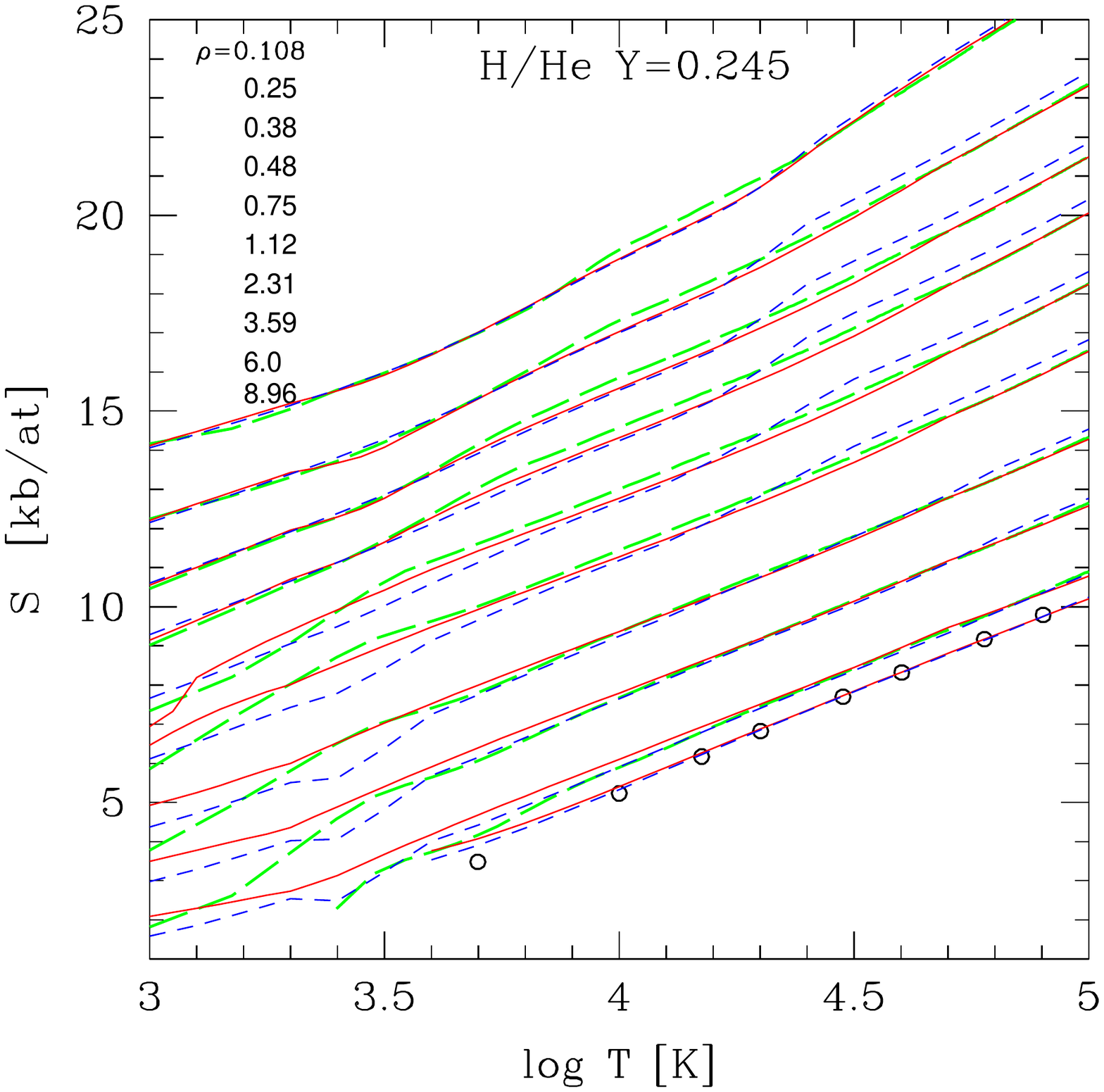}
\vspace{-2cm}
\caption{Entropy vs temperature for several isochore calculations by Militzer \& Hubbard (2013) (as labeled in the figure), compared with the present and SCvH results, respectively. Same labeling as in the previous figures. Careful: for sake of clarity, each curve from $\rho=3.59\gcc$ to $\rho=0.108\gcc$ has been shifted upward by 1 $k_b/$atom w.r.t. to the
immediately higher-density one.} 
\label{fig26}
\end{figure}

Figures \ref{fig25}  and \ref{fig26} show the same comparisons for the entropy. 
For the coolest isotherms ($T<5000$ K) and low densities
($\rho \lesssim 0.3\gcc$), i.e. in the molecular/atomic domain, all calculations agree quite well, showing that the SC model adequately handles this regime, even when interactions
between H$_2$ molecules or He atoms become significant. For higher temperatures and densities, the SC model starts to depart from both the MH13 and present results, first {\it underestimating} the entropy in the domain $0.2\lesssim \rho \lesssim 2\gcc$ and $5000\lesssim T\lesssim$ 10000 K and then showing an abrupt
{\it increase} of the entropy in this density regime  at higher temperatures. This corresponds to the very domain of pressure ionization and reflects the already mentioned
inaccurate (and too abrupt) treatment of this process in the SC theory. In contrast, the agreement between the present calculations and the MH13 results can be considered as very satisfactory over the
entire temperature-density range explored by the simulations. The sudden decrease in entropy for $T=5000$ K and $\rho\gtrsim 2\gcc$ in the fit derived from the MH13 simulations compared with both the
present and SCvH results, as seen  in both Figures \ref{fig25} and \ref{fig26}, 
is rather surprising and might point to either the increasing contribution to the interactions between H and He species or an issue with the TDI procedure or the inferred fitting formula. 
Note, however, that quantum effects between ions become significant in this regime (see Fig. \ref{fig1}) and that either
the present calculations treat them as a perturbation, with the Wigner-Kirkwood expansion (SCvH and present) or they are ignored (MH13). 
As seen in the figures, however, this region concerns a high density and low entropy ($S<6\,k_b/$e$^-$) domain where there are
 no astrophysical objects. 

\begin{figure}
\includegraphics[width=\linewidth]{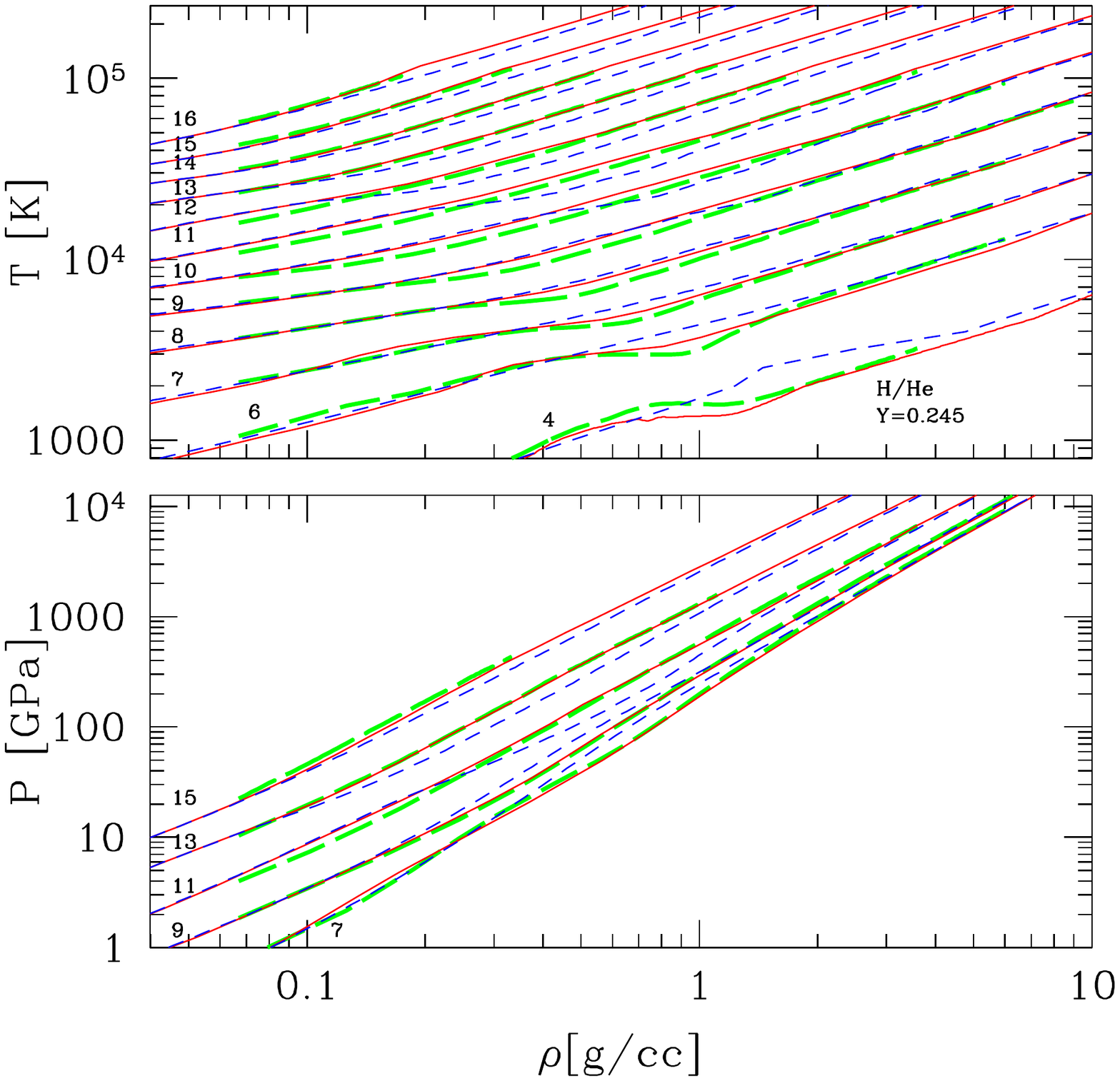}
\vspace{-2cm}
\caption{Temperature and pressure profiles for a series of adiabats as labeled in the figure in $\kb/e^-$ ($=(\kb/{\rm atom})/1.076$ for the present $Y$ value) for the MH13 mass fraction
of helium ($Y=0.245$). Solid lines: present calculations; green long-dashed lines: MH13; blue short-dashed lines: SCvH.} 
\label{fig27}
\end{figure}

As mentioned above, interiors of astrophysical bodies below $\sim 0.6\,\msol$ are essentially convective and thus nearly adiabatic. Their internal profile is thus characterized by an
isentrope for a given mass at a given age and their thermal evolution corresponds to a series of decreasing isentropes. Figure \ref{fig27} portrays the temperature and pressure
profiles of such isentropic structures for the present, SCvH and MH13 calculations for entropy values between 4 and 16 $k_b/$e$^-$, i.e. $2.9\times10^{-2}$ to  $1.15\times10^{-1}$ MJ kg$^{-1}$ K$^{-1}$ for the present H/He mixture ($Y=0.246$, i.e. $x_{He}=0.076$). 
For adiabats above $S\gtrsim 10\,k_b/$e$^-$, we note that the SCvH
adiabats underestimate both the temperature and the pressure in the density domain $\sim 0.1$-2.0 $\gcc$, i.e. the domain of ionization,
compared with the two other types of calculations, both based on QMD calculations in this regime, which predict 
higher temperatures and pressures in this domain. Again, this reflects the inaccurate treatment of the ionization process (Stark effect and electronic excitations) in the SC model.

Interestingly enough, the behaviur
reverses for cooler isentropes, with SCvH predicting higher temperatures and pressures than the two other models. In this regime, molecular hydrogen H$_2$ is still present and the
disagreement arises from the lack of a proper treatment of H$_2$ pressure dissociation in the SCvH model. 
As seen in the figure, the present EOS agrees fairly well with the calculations of MH13, notably for the pressure. 
For the temperature, the present calculations start departing from MH13
for $S\lesssim 10\,k_b/$e$^-$, predicting slightly warmer structures than these latter calculations in the pressure ionization regime. 
This reflects the increasing contribution of the H/He interactions, and thus of the non-ideal mixing entropy, in the
mixture, not treated in the present calculations,
for the cooler and denser domains, yielding eventually a H/He phase separation for the coolest isentropes (Lorenzen et al. 2009, Morales et al. 2010b, Militzer 2013, Mazzola et al. 2018, Sch\"ottler \& Redmer 2018).

Since we do not have the results from MH13 for the exact Jupiter isentrope ($T=166$ K, $P=1$ bar) 
for the cosmogonic helium mass fraction ($Y=0.275$), we can not make a comparison for the correct Jupiter adiabat. 
The $S=7\,k_b/$e$^-$ one, however, is close enough to Jupiter's value to estimate the discrepancy between the various calculations under Jupiter-like conditions. As can be inferred from
Fig. \ref{fig25}  and \ref{fig26}, we found out that, for this
entropy value, the maximum discrepancy occurs at $P=100$ GPa, with MH13 giving a temperature of $T=4867$ K, the present EOS $T=5382$ K, i.e. a 10$\%$ difference, and SCvH $T=5623$ K, i.e. 15.5$\%$ difference. 
Conversely, for the corresponding density $\rho\simeq 0.75\gcc$, MH13 predicts a pressure $P=94$ GPa against 93 GPa for the present calculations ($\approx 1\%$ discrepancy) and 132 GPa for SCvH ($=37\%$ discrepancy). 
This is the very domain of pressure ionization, so these differences are not surprising
and illustrate the better treatment of this process in the present calculations, based on QMD
simulations in this domain, compared with SCvH. 
The remaining discrepancy with MH13 can thus have two origins. The first one is errors in the parameterization of the
free energy in the present calculations.
The second one is of course
the missing treatment of H/He interactions in the present EOS calculations and thus the lack of non-ideal mixing entropy. Indeed, MH13 numerical simulations reveal an H/He phase separation in this regime
(see Fig. 2 of Militzer 2013), even though other simulations reach a different conclusion for similar
$T$-$P$ values (Sch\"ottler \& Redmer 2018), suggesting that H and He are still miscible under
Jupiter internal adiabat conditions. Looking at figures \ref{fig25}, \ref{fig26} and \ref{fig27}, 
we can infer the impact of the aforementioned discrepancies in $T$ and $P$ between the present, SCvH and MH13 calculations 
in the pressure ionization
domain under Jupiter-like conditions.
We found out that for $T=5000$ K and $\rho=0.75\gcc$ (and $Y=0.246$), we get $S=6.9$ and 6.63 $\kb$/e$^-$ for the present and SCvH EOS,
respectively, against 7.07 $\kb$/e$^-$ for MH13, i.e. $\sim$2\% and 6\% differences, respectively.
A precise quantification of these differences upon Jupiter's internal properties requires deeper explorations, with exact models of Jupiter, to
be conducted in forthcoming calculations. 

Note that, in the above comparisons, we have not included the EOS recently derived by Miguel et al. (2016+erratum). Indeed, the entropy values
given by their tables for various $T$ and $\rho$ conditions differ significantly from {\it all} the calculations displayed in this section.
This points to a severe issue in these tables (see also Debras \& Chabrier 2019, \S2.2).

\subsection{Thermodynamic quantities}

As mentioned earlier,
the present EOS delivers all the necessary thermodynamic quantities besides temperature $T $, pressure $P$, specific internal energy $\ubar$ and specific entropy $\sbar$. These include the specific heats at constant volume and pressure, $\ucv$, $\ucp$ (see eqn.(\ref{cpcv})), and the adiabatic gradient, $\gad$, from the relation

\begin{eqnarray}
\gad &=& (\frac{\partial \log T}{\partial \log P})_S=
	- { ({\partial \log S\over \partial \log P})_T \over ({\partial \log S \over \partial \log T})_P }
	= {\chi_T\over {\chi_T^2+\chi_\rho {\ucv \over P/\rho T} } },
\label{gad}
\end{eqnarray}
where $\chi_T=(\frac{\partial \log P}{\partial \log T})_\rho=-(\frac{\partial \log \rho}{\partial \log T})_P/
(\frac{\partial \log \rho}{\partial \log P})_T$ and $\chi_\rho=(\frac{\partial \log P}{\partial \log \rho})_T$.
Note that since, in our calculations, the ideal mixing entropy, $S_{mix}$, does not depend on $T$ or $P$ (see eq.(\ref{Sid})), the adiabatic gradient (as well as the other first derivative quantities) at given $(P,T$) for the mixture can easily be calculated from the linear interpolation
\begin{eqnarray}
\gad(X,P,T) = 
	 {\sum_i X_i S_i ({\partial \log S_i\over \partial \log P})_T  \over \sum_i X_i S_i ({\partial \log S_i \over \partial \log T})_P },
\label{gad}
\end{eqnarray}
which may happen to be less numerically noisy than calculating the second derivative of a spline. Other quantities include the thermal expansion coefficient $\alpha$, the adiabatic sound speed $C_S$, the isothermal and isentropic compressibilities $\kappa_T, \kappa_S$, and the Gr\"uneisen parameter $\gamma$, all easily derived from these relations:

\begin{eqnarray}
\alpha&=&-\frac{1}{ T}(\frac{\partial \log \rho}{ \partial \log T})_P=+\frac{1}{ T}{\chi_T \over \chi_\rho}\\
C_S &=& (\frac{\partial P}{ \partial \rho})_S^{1/2}=(\frac{P}{\rho})^{1/2}(\frac{C_P}{ C_V})^{1/2}\chi_\rho^{1/2}\nonumber \\
       &=&\bigl\{ {\rho \over \chi_\rho P} (1-\chi_T \gad) \bigr\}^{-1/2}\\
\frac{\kappa_T}{\kappa_T^0}&=&({\rho \kb T\over \mu P})\chi_\rho^{-1}\\
\kappa_S&=&{1\over \rho C_S^2}\\
\gamma&=&\frac{1}{ \rho} (\frac{\partial P}{ \partial U})_\rho={\alpha \over \rho C_V\kappa_T}={P\over \rho T C_V}\chi_T,
\label{speed}
\end{eqnarray}
where $\mu=\langle A\rangle \mh$ and $\kappa_T^0=({\rho \kb T\over \mu })^{-1}$ denotes the value for a perfect gas of atomic mass $\mu$.

\begin{figure}
\includegraphics[width=\linewidth]{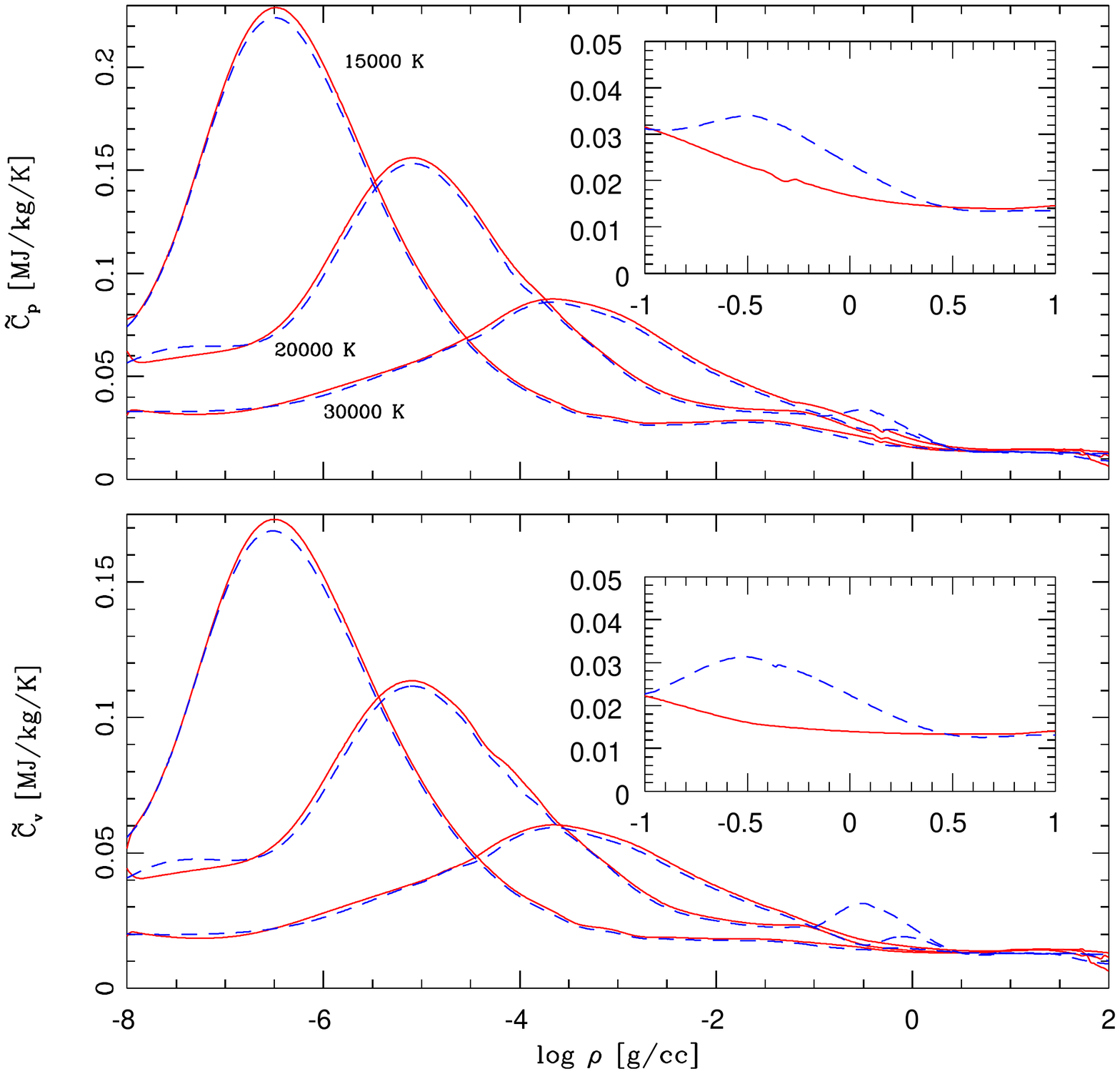}
\vspace{-2cm}
\caption{Specific heats at constant pressure and constant volume as a function of density for 3 isotherms (as labeled in the figure) for
a cosmogonic helium abundance ($Y=0.275$).
The inset highlights the pressure dissociation/ionization domain for $T=20000$ K.  Solid red: present; blue long-dashed: SCvH.} 
\label{fig28}
\end{figure}

\noindent Figure \ref{fig28} portrays the specific heats for an H/He mixture with solar helium abundance ($Y=0.275$). We recover the same features as in Fig. \ref{fig14} with the obvious domains of H$_2$ temperature and pressure dissociation and ionization.

\begin{figure}
\includegraphics[width=\linewidth]{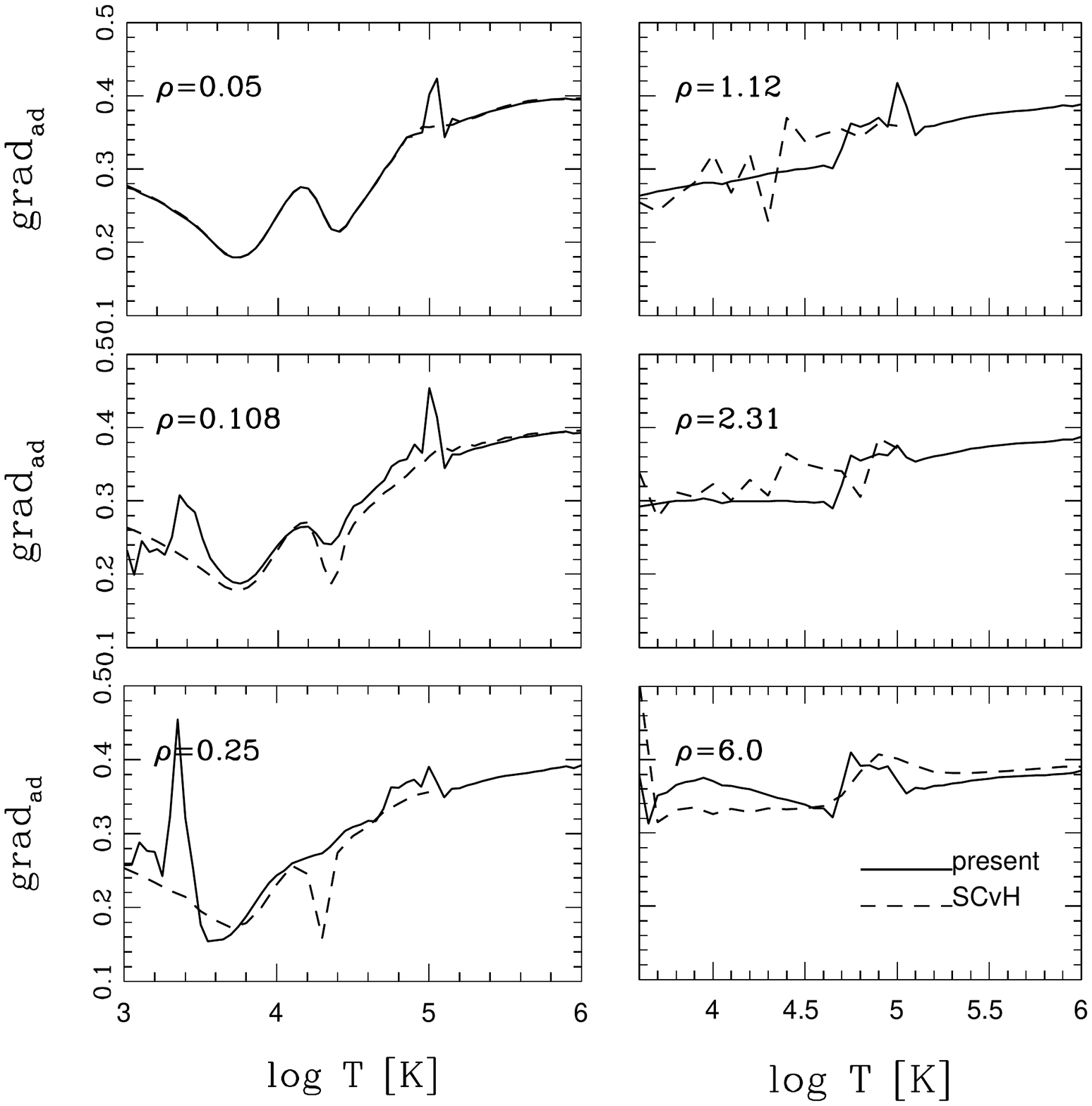}
\vspace{-2cm}
\caption{Adiabatic gradient as a function of temperature for various isochores for
a cosmogonic helium abundance ($Y=0.275$).
Density (marked in $\gcc$) increases from top to bottom and left to right. Notice the change of
scale in the temperature axis ($x$-axis) in the right panel.}
\label{fig29}
\end{figure}

The adiabatic temperature gradient is a quantity of prime importance in astrophysics because it is the quantity used in the Schwarzschild criterion to
determine whether transport of energy occurs by convection or by microscopic
diffusion processes (conduction or radiation). 
This quantity is displayed in Fig. \ref{fig29} for the same cosmogonic H/He abundance ($Y=0.275$). 
The figure spans a $\rho$-$T$ range characteristic of
pressure dissociation and ionization from 10$^3$ to 10$^6$ K and 0.05 to 6.0 $\gcc$.
Some typical physical features can be identified in the figure. At low density,
the low temperature limit corresponds to the domain of 
molecular hydrogen with excited rotational levels ($\theta_{rot}\le T<
\theta_{vib}$), $\gad=0.3$, while the high temperature domains corresponds to the 
perfect monatomic gas, $\gad=0.4$, potentially modified by non-ideal contributions.
The two dips reflect the excitation of vibrational levels and the regions of hydrogen dissociation and H or He ionization, respectively. 
As density increases and dissociation/ionization take place, the two dips vanish and eventually the whole mixture becomes fully ionized for $\rho \ga 6\gcc$. The spikes around $\log T\approx 3.4$ for $\rho=0.108$ and 0.25 $\gcc$ reflect the onset of hydrogen crystallization.
Being a combination of several second derivatives, the adiabatic gradient is very sensitive to the interpolation procedures in the calculations
of the EOS table. This is reflected
 by the wiggles in the domain $4.7 \lesssim \log T \lesssim 5.0$, which is the domain of interpolation between
the QMD and CP98 calculations. This is the unfortunate consequence of the necessity to combine different calculations in order to construct large enough
$(T,P,\rho)$ tables for astrophysical use.

\section{Form of the EOS tables}

As mentioned in the Introduction, QMD calculations for the EOS have been performed in the canonical ensemble, i.e. with ($T,\rho$) as
independent variables. The additive volume law (AVL) procedure
to calculate the thermodynamic quantities of the H/He mixture, however, requires the independent variables to be ($T,P$), which imposes
transformations of the various quantities from the ($T,\rho$) ensemble into the ($T,P$) ensemble
 by spline interpolation procedures. Eventually, the ($T,P$) table for the H/He mixture was
transformed back into a ($T,\rho$) one, as many astrophysical calculations use these latter quantities as input variables.
Although the online H/He table corresponds to a solar (cosmogonic) helium abundance, $Y=0.275$, other mixtures can be easily obtained
from the pure H and He tables with use of eqns.(\ref{W})-(\ref{Sid}).

For reasons of practical interpolation, all tables have rectangular forms with the following limits:

\begin{eqnarray}
2.0 &\le& \log\, T\le 8.0,Ê\nonumber \\
 -9.0&\le& \log\, P\le +13.0, \nonumber \\
 -8.0 &\le& \log\, \rho \le +6.0,
\end{eqnarray}
with grid spacings $\Delta \, \log T=0.05$, $\Delta \, \log P=0.05$, $\Delta \, \log \rho=0.05$, i.e.  121 isotherms, each with 441 values of $P$ or 281 values of $\rho$, and $T$ in K, $P$ in GPa, $\rho$ in $\gcc$. As 
mentioned in the previous sections, some parts of these tables are unphysical because they correspond to regions in the
diagrams that are not handled by the present calculations. These regions concern essentially the domains of solid hydrogen and helium
and regions where quantum diffraction effects on the nuclei can no longer be treated by a Wigner-Lirkwood expansion.
They are identified in Fig. \ref{fig1} and \ref{fig16}. 
Because of the rectangular format of the tables, values at very low density in the $(\log T, \log P$) table and at very low pressure in the $(\log T, \log \rho)$ tables also become unphysical and should not be considered.

Table I gives an example of the various quantities provided by the tables. All quantities are specific quantities, i.e. are given by unit mass,
with $A_H=1.00794$, $A_{He}=4.00262$ and the atomic mass unit $m_H=1.66\times 10^{-27}$ kg.
The main second derivatives are also provided. All necessary thermodynamic quantities can be derived from these derivatives, from eqns(\ref{cpcv}) and (\ref{gad})-({\ref{speed}).
Users, however, might prefer to use only values corresponding to the
first derivatives of the Helmholtz free energy, namely $U$, $P$, $S$, and to carry out their own interpolation procedures to calculate second
derivatives. }

We also stress that the entropy of the spin of the nuclei $S_{nuc}^{id}=\ln(2s+1)$ $\kb$/proton, where $s$ is the spin of the nucleus, is {\it not} included in the calculations.

\begin{table*}
\centering
\begin{tabular}{*{10}{c}} 
\hline
\hline
\\
 $\log\,T$ & $\log\,P$ & $\log\,\rho$ & $\log\,{\tilde U}$ & $\log\,{\tilde S}$ & $({\partial \log \rho\over \partial \log T})_P$ & $({\partial \log \rho\over \partial \log P})_T$ & $({\partial \log S\over \partial \log T})_P$ & $({\partial \log S\over \partial \log P})_T$ & $\nabla_{ad}$ \\\\ 
\hline
\\
\#iT=1 log T= 2.000 \\
      0.200E+01 & -0.900E+01 & -0.909E+01 & -0.628E-01 & -0.100E+01 &  -0.114E+01 &  0.983E+01 &  0.242E+02 &  -0.355E+00  & 0.300E+00 \\
 0.200E+01 & -0.895E+01  &  -0.870E+01  & -0.628E-01  &  -0.102E+01   &  -0.591E+01 &  0.572E+01 &  0.190E+02 &  -0.297E+00  & 0.300E+00   \\
    ...  &  ...   & ...    &   ...  &  ... &  ...    &   ...   &  ...  &    ...  & ...    \\ 
       ...  &  ...   & ...    &   ...  &  ... &  ...    &   ...   &  ...  &    ... & ...       \\ 
   0.200E+01  &    0.129E+02  &    0.542E+01   &   0.781E+01   &   0.547E+01   &  0.454E+00   &   0.500E+01   &  -0.152E+01   &    0.506E+00       & 0.330E+00 \\ 
   0.200E+01   &   0.130E+02   &   0.544E+01   &   0.784E+01 &     0.550E+01 &    0.455E+00 &     0.498E+01   &   -0.152E+01 &     0.506E+00     &  0.330E+00 \\ 
   0.205E+01  -&  -0.900E+01  &   -0.978E+01  &   -0.174E-01   &  -0.402E+00   &   -0.160E+02    &  0.161E+02  &    -0.260E+01    &  -0.315E+01     & 0.300E+00  \\ 
   0.205E+01  &   -0.895E+01  &   -0.911E+01  &   -0.174E-01  &   -0.548E+00  &    -0.105E+02   &   0.104E+02   &   -0.168E+01  &   -0.267E+01     & 0.300E+00  \\ 
    ...  &  ...   & ...    &   ...  &  ... &  ...    &   ...   &  ...  &    ... &  ...    \\ 
       ...  &  ...   & ...    &   ...  &  ... &  ...    &   ...   &  ...  &    ... & ...      \\ 
\#iT=2 log T= 2.050 \\
0.205E+01   &   -0.900E+01  &   ... & ... \\
    ...  &  ...   & ...    &   ...  &  ... &  ...    &   ...   &  ...  &    ... &  ...    \\ 
       ...  &  ...   & ...    &   ...  &  ... &  ...    &   ...   &  ...  &    ... & ...      \\ 
   0.205E+01   &   0.130E+02    &  0.547E+01   &   0.781E+01   &   0.542E+01   &  0.451E+00    &  0.498E+00   &   -0.152E+01   &   0.507E+00       & 0.331E+00  \\ 
 \#iT=3 log T= 2.100 \\
   0.210E+01   &   ...  &   ...  \\
    ...  &  ...   & ...    &   ...  &  ... &  ...    &   ...   &  ...  &    ...  &  ...      \\ 
       ...  &  ...   & ...    &   ...  &  ... &  ...    &   ...   &  ...  &    ... & ...      \\ \\
\hline
\end{tabular}  
\caption{Example of the EOS table. Units are $T$ in K, $P$ in GPa, ${\tilde U}$ in MJ kg$^{-1}$, ${\tilde S}$ in MJ kg$^{-1}$ K$^{-1}$.
Each ($\log T, \log P$) or  ($\log T, \log \rho$) table includes $N_T=121$ isotherms, each with $N_P=441$ pressure values or $N_\rho=281$ density values,
with step values 0.05 for $\Delta \log T$, $\Delta \log \rho$ and $\Delta \log P$.
Note: the number of digits after the point has been truncated in this exemple to fit the journal format. In the online table, all quantities are given with 6 digits after the point.
}
\label{Table_eos}
\end{table*}

\section{Conclusion}

In this paper, we have presented new equations of state for pure fluid hydrogen and helium as well as for hydrogen/helium mixtures within
the so-called additive volume law (AVL) approximation, i.e. simply taking into account the ideal mixing entropy contribution between the two species (H and He) to the thermodynamic
quantities of the mixture. The calculations combine first-principle calculations, based on quantum molecular dynamics (MD-DFT)
simulations, in the regime of pressure ionization, with semi-analytical calculations in the low density (molecular/atomic) and high density or temperature (fully 
ionized) regimes, and provide not only the pressure, internal energy and density but also the entropy and
all necessary thermodynamic derivatives. 
The initial calculations are performed in the canonical ensemble, implying ($T$, $\rho$) as independent variables, and are transformed into ($T,P$)
tables to be able to make use of the AVL for the mixture. Therefore, we provide tables in both sets of independent variables.
The EOS tables cover a wide temperature-pressure-density domain which permits the calculations of 
the mechanical and thermal (cooling) structures of a large variety of astrophysical
bodies, from solar-type stars to low-mass stars, brown dwarfs and (solar and extrasolar) gaseous planets down to Saturn-like masses. 
These calculations should supersede the previously widely used Saumon-Chabrier-vanHorn EOS in this domain. At higher densities and/or
temperatures, the EOSs merge with those of Chabrier \& Potekhin (1998) and Potekhin \& Chabrier (2000) ones, devoted to the physics of compact,
relativistic bodies such as white dwarfs and neutron stars.

These calculations are by no means without flaws and limitations. Flaws include unphysical numerical oscillations, notably in the calculations of second-derivative
thermodynamic quantities, due to spline interpolations. For this reason, verifications of the Maxwell
relations between thermodynamic derivatives would be meaningless, because they would undoubtedly be affected by the
numerical interpolation procedures and thus have no real physical foundations (see, e.g. \S8 and
Figs. 18 and 19 of SCvH). 
Possible future improvements in these numerical treatments will be indicated in future versions of the online EOS tables. 
Note also that QMD simulations retain as well some degree of uncertainty, inherent to
the exchange-correlation functional used in the calculations. Indeed, it has been shown that for liquid hydrogen, pressures obtained with the Perdew-Burke-Ernzerhof functional (Perdew et al. 1996) and by the van der Waals functional (Lee et al. 2010) functionals can differ by as much as $\sim 10$-20\% for a given density in the present domain of interest (Morales et al. 2013; see also Pierleoni et al. 2016, Knudson \& Desjarlais 2017,  Mazzola et al. 2018).

The most challenging limitation of the present calculations is the use of the AVL in the treatment of the H/He mixture, which omits interactions
between hydrogen and helium species. While relatively inconsequential for bodies with internal entropies larger than about 10 $\kb/e^-$ ($\sim 11\,\kb/at$ ), i.e. about $7\times 10^{-2}$ MJ kg$^{-1}$ K$^{-1}$, as seen from the comparisons with Militzer \& Hubbard (2013)
(see Fig. \ref{fig27}), this contribution becomes important for cooler entropy values, which encompasses essentially all
objects in the brown dwarf and planetary domain older than a few gigayears.
Incorporating these non-ideal H/He contributions into the present calculations will be explored in the near future.

The H, He and H/He EOS tables for a solar mixture ($Y=0.275$) are available on the site: http://perso.ens-lyon.fr/gilles.chabrier/DirEOS

\acknowledgments
The authors are deeply indebted to Andreas Becker and Ronald Redmer for providing tables of their EOS
 calculations and to Burkhard Militzer for making available
his own calculations, which enabled us to perform detailed comparisons between the different EOSs. We also thank St\'ephanie Brygoo, Paul Loubeyre and Markus
Knudson for sending us their data. This work has been partly supported by the Programme National de Plan\'etologie (PNP).
\\

\centerline{REFERENCES}
\noindent 
Baraffe, I., Chabrier, G., Allard, F., \& Hauschildt, P., 2003, \aap, 402, 701 \\
Baraffe, I., Homeier, D., Allard, F., \& Chabrier, G., 2015, \aap, 577, 42\\
Becker, A., Nettlelmann, N., Holst, B., \& Redmer, R., 2013, \prb, 88, 45122 \\
Becker, A., et al., 2014, \apjs, 215, 21 \\
Belov, S., et al., 2002, JETP Lett., 76, 443 \\
Bonev, S., Schwegler, E., Ogitsu, T., \& Galli, G., 2004, {\it Nature}, 431, 669 \\
Boriskov, G.V., et al., 2003, Dokl. Phys., 48, 553   \\
Boriskov, G.V., et al., 2011, CoPP, 51, 339   \\
Brygoo, S.  et al., 2015, J. Appl. Phys., 118, 195901 \\
Caillabet, L., Mazevet, S., \& Loubeyre, P., 2011, \prb, 83, 4101    \\
Celliers, P., et al., 2010, \prl, 104, 184503 \\
Chabrier, G., 1990, J. de Physique, 51, 1607   \\
Chabrier, G., 1993, \apj, 414, 695 \\
Chabrier, G., \& Potekhin, A., 1998, \pra, 58, 4941   \\
Chabrier, G., \& Ashcroft, N., 1990, \pra, 42, 2284   \\
Chabrier, G. \& Baraffe, I., 2000, \araa, 38, 337  \\
Collins, G.W., et al., 1998, Science, 281, 1178 \\
Datchi, F., Loubeyre, P., \& Le Toullec, R., 2000, \prb, 61, 6535 \\
Debras, F. \& Chabrier, G., 2019, \apj, in press \\
Deemyad, S., \& Silvera, I., 2008, \prl, 100, 155701 \\
Eremets, M., \& Trojan, I., 2009, {\it ZhETF} 89, 198 \\
Fortov, V., et al., 2007, \prl, 99, 185001 \\
Hicks, D.G., et al., 2009, \prb 79, 014112\\   
Holst, B., Redmer, R., \& Desjarlais, M., 2008, 77, 184201 \\
Hu, S.X., Militzer, B., Goncharov, V., \& Skupsky, S., 2011, \prb, 84, 224109  \\
Kechin, V., J. Phys.: Condens. Matter, 7, 531 \\
 Knudson, M., 2004, \prb, 69, 4209   \\
Knudson, M. \& Desjarlais, M., 2009, \prl, 103, 225501 \\
Knudson, M., \& Desjarlais, M., 2017, \prl, 118, 035501  \\
Lee, K., Murray, E., Kong, B., Lundqvist, B., \& Langreth, D., 2010, \prb, 82, 081101 \\
Loubeyre, P., Le Toullec, R. \& Pinceaux, J-P., 1993, \prl, 70, 2106  \\
Loubeyre, P. et al., 2012, \prb, 86, 144115  \\
Lorenzen, W., Holst, B., \& Redmer, R., 2009, \prl, 102, 115701  \\
Lorenzen, W., Holst, B., \& Redmer, R., 2011, \prb, 84, 235109  \\
Mazzola, G., Helled, R., \& Sorella, S., 2018, \prl, 120, 5701   \\
Miguel, Y., Guillot, T., \& Fayon, L., 2016, \aap, 596, 114  \\
Militzer, B., 2009, \prb, 79, 5105   \\ 
Militzer, B., 2013, \prb, 87, 4202   \\
Militzer, B., \& Ceperley, D., 2001, \pra, 63, 6404  \\  
Militzer, B., \& Ceperley, D., 2000, \prl, 85, 1890 \\   
Militzer, B., \& Hubbard, W., 2013, \apj, 774, 148   \\
Mochalov, M. A., 2010, JETP Lett., 2010, 92, 300 \\
Morales, M., Pierleoni, C., \& Ceperley, D., 2010a, \pre, 81, 021202 \\
Morales, M., Pierleoni, C., Schwegler, \& Ceperley, D., 2010b, PNAS, 107, 12799 \\
Morales, M., McMahon, J., Pierleoni, C., \& Ceperley, D., 2013, \prl, 110, 065702 \\
Pierleoni, C., Morales, M., Rillo, J., Holtzmann, M., \& Ceperley, D., 2016, PNAS, 113, 4953 \\
Perdew, J., Burke, K. \& Ernzerhof, M., 1996, \prl, 77, 3865 \\
Potekhin, A., \& Chabrier, G., 2000, \pra, 62, 8554   \\
Sano, T., et al., 2011, \prb 83, 54117   \\
Saumon, D. \& Chabrier, G., 1991, \pra, 44, 5122   \\
Saumon, D. \& Chabrier, G., 1992, \pra, 46, 2084 \\ 
Saumon, D., Chabrier,G., \& van Horn, H., 1995, \apjs, 99, 713  \\ 
Sch\"{o}ttler, \& Redmer, R., 2018, \prl, 120, 5703  \\
Soubiran, F., Mazevet, S., Winisdoerffer, C., \& Chabrier, G., 2012, \prb, 86, 115102   \\
Soubiran, F., Mazevet, S., Winisdoerffer, C., \& Chabrier, G., 2013, \prb, 87, 165114   \\
Soubiran, F. \& Militzer, B., 2013, \apj, 806, 228   \\
Wigner, E., \& Huntington, H., 1935, \jcp, 3, 764    \\
Winisdoerffer, C., \& Chabrier, G., 2005, \pre, 71, 026402  \\ 
Zel'dovich \& Raizer, 2002, Physics of Shock Waves and High Temperature Hydrodynamic Phenomena, Dover   

\end{document}